%% file: main.tex
\documentclass[lettersize,journal]{IEEEtran}

\IEEEoverridecommandlockouts                              
                                                          
\usepackage{amsmath,amsthm}
\usepackage[utf8]{inputenc} 
\usepackage[T1]{fontenc}    
\usepackage{url}            
\usepackage{booktabs}       
\usepackage{amsfonts}       
\usepackage{nicefrac}       
\usepackage{microtype}      
\usepackage{xcolor}         
\usepackage{graphicx}
\usepackage{lipsum}
\usepackage{bm}
\usepackage{mathtools}
\usepackage[normalem]{ulem}
\usepackage{cuted}
\usepackage[splitrule]{footmisc}
\usepackage{algorithm}
\usepackage{algpseudocode}
\usepackage{array}
\usepackage{caption}
\usepackage{cite}
\usepackage{caption}
\usepackage{subcaption}
\usepackage[colorlinks,bookmarks=false]{hyperref}
\hypersetup{citecolor=[RGB]{0,0,255}, urlcolor=[RGB]{0,0,255}}
\usepackage{multicol}
\usepackage{multirow}

\newtheorem{definition}{Definition}

\newtheorem{theorem}{Theorem}
\newtheorem{lemma}{Lemma}
\newtheorem{corollary}{Corollary}
\newtheorem{remark}{Remark}

\newcommand{\revise}[1]{\textcolor{black}{#1}}

\newcommand{\diag}{\text{diag}}
\newcommand{\sip}{\text{sp}}
\usepackage{mathtools}
\DeclarePairedDelimiter{\ceil}{\lceil}{\rceil}
\newcommand*{\bomega}[0]{\boldsymbol{\omega}}
\newcommand*{\bdelta}[0]{\boldsymbol{\delta}}

\renewcommand{\baselinestretch}{.99}

\newcommand\oprocendsymbol{\hbox{$\bullet$}}
\newcommand\oprocend{\relax\ifmmode\else\unskip\hfill\fi\oprocendsymbol}

\usepackage{amsmath}
\usepackage{array}
\newcommand{\real}[0]{\mathbb R}

\begin{document}
\title{Online Event-Triggered Switching for
Frequency Control in Power Grids with Variable Inertia}

\author{Jie Feng~\IEEEmembership{Student Member,~IEEE}, Wenqi Cui~\IEEEmembership{Student Member,~IEEE}, \\ Jorge Cort\'es~\IEEEmembership{Fellow,~IEEE}, Yuanyuan Shi~\IEEEmembership{Member,~IEEE}
\thanks{This work was supported under grants NSF ECCS-2200692, NSF ECCS-1947050, and Jacobs School Early Career Faculty Development Award 2023. Jie Feng is supported by UC-National Laboratory In-Residence Graduate Fellowship L24GF7923. Yuanyuan Shi is partially supported by the Hellman Fellowship. Jie Feng and Yuanyuan Shi are with the Department of Electrical and Computer Engineering, UC San Diego. Wenqi Cui is with the Department of Electrical and Computer Engineering, University of Washington, Seattle. Jorge Cort\'es is with the Department of Mechanical and Aerospace Engineering, UC San Diego.}
}

\maketitle

\begin{abstract}
The increasing integration of renewable energy resources into power grids has led to time-varying system inertia and consequent degradation in frequency dynamics. A promising solution to alleviate performance degradation is using power electronics interfaced energy resources, such as renewable generators and battery energy storage for primary frequency control, by adjusting their power output set-points in response to frequency deviations. However, designing a frequency controller under \emph{time-varying inertia} is challenging. \revise{Specifically, the stability or optimality of controllers designed for time-invariant systems can be compromised once applied to a time-varying system.}
We model the frequency dynamics under time-varying inertia as a nonlinear switching system, where the frequency dynamics under each mode are described by the nonlinear swing equations and different modes represent different inertia levels. 
We identify a key controller structure, named Neural Proportional-Integral (Neural-PI) controller, that guarantees exponential input-to-state stability for each mode. 
To further improve performance, we present an online event-triggered switching algorithm to select the most suitable controller from a set of Neural-PI controllers, each optimized for specific inertia levels. Simulations on the IEEE 39-bus system validate the effectiveness of the proposed online switching control method with stability guarantees and optimized performance for frequency control under time-varying inertia.
\end{abstract}


\IEEEpeerreviewmaketitle
\section{Introduction}
Frequency stability is vital for the security and operation of power systems, the goal of which is to balance power generation and demand to maintain the system frequency near its nominal value (i.e., 60 Hz in the US). This paper mainly focuses on primary frequency control, which corrects immediate power imbalances within seconds \cite{kundur1994power}. 
The surge in integrating renewable energy sources like wind and solar, while marking significant progress towards sustainability, introduces larger fluctuations in net loads due to their unpredictable power outputs, thus requiring more advanced controllers \cite{9721402}. Moreover, many of these new technologies are interfaced with the grid through power electronic interfaces (i.e., inverters), which have no rotational inertia. At the same time, the grid still has a large number of synchronous components, creating a system that is a mixture of conventional machines and inverter-connected resources. The amount of inertia depends on the amount of online synchronous generators. \revise{At different times of the day, renewable generations will displace a different amount of power generation from synchronous machines, leading to different numbers of online synchronous generators. As a result, the grid can present a \emph{reduced} and \emph{switching} system inertia, where switching means the inertia is a right continuous piecewise constant function of time~\cite{nrel_inertia_report}.}
The resulting complexity has been linked to a noticeable degradation in system frequency dynamics \cite{ULBIG20147290,7866938,9683585}, risking load shedding and blackouts. \revise{Despite significant efforts for handling reduced inertia \cite{8226839,9107487,9721402}, two facts motivate further research considering the time-varying systems: (1) even if each sub-system is exponentially stable, the switching system can be unstable \cite{VU2007639}; (2) the optimal controller for a specific inertia can be suboptimal for another. For example, as system inertia increases, frequency dynamics slow down, potentially causing prolonged frequency deviations with the controller optimized for low inertia~\cite{9161355}. }

\begin{figure}[t]
    \centering
    \includegraphics[width=0.5\textwidth]{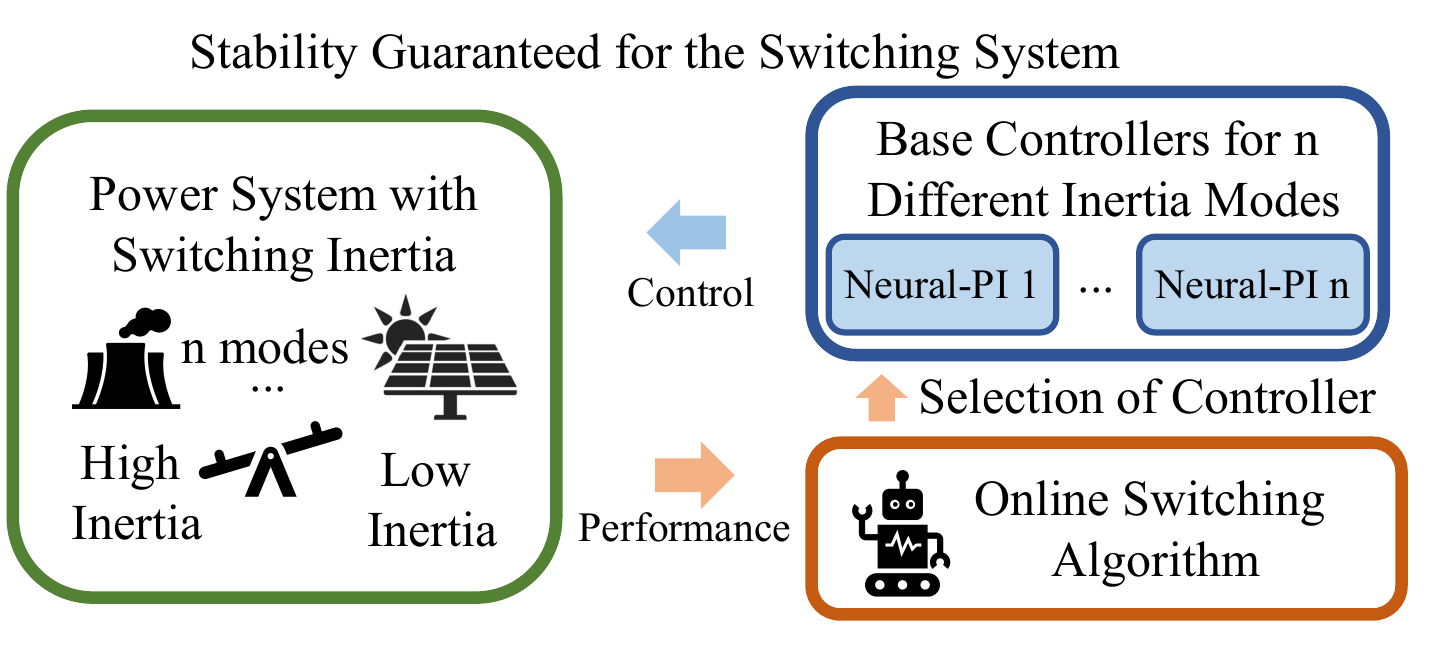}
    \caption{The proposed online switching control method for frequency control under variable inertia with stability guarantee. \vspace{-0.5cm}}
    \label{fig:switching-diagram}
\end{figure}

To tackle the challenge of frequency control under time-varying inertia, we model the system as a nonlinear switching system between 
different inertia levels, where the switching signal is \emph{unknown} to the controller beforehand. \revise{Our objectives are first to guarantee stability for the time-varying system and then to improve the performance of controllers designed for time-invariant systems.} We first identify a key controller structure, named Neural Proportional-Integral (Neural-PI) control, that guarantees exponential input-to-state stability (Exp-ISS) of the nonlinear frequency dynamics under an arbitrary, fixed inertia mode, through Lyapunov analysis. The Neural-PI structure is first introduced in~\cite{cui2023structured} for frequency control under time-invariant inertia with asymptotic stability guarantees. In this work, we first improve the analysis of~\cite{cui2023structured} to show the Neural-PI controller guarantees Exp-ISS for the time-invariant system, and then extend the Exp-ISS guarantees of the Neural-PI controller to nonlinear switching systems.
To further improve the controller performance \revise{for the time-varying system}, we propose an online event-triggered switching algorithm for \emph{dynamic controller selection} from a set of pre-trained Neural-PI controllers, each of which is trained under a specific inertia level, and the switching algorithm online updates the selection probability according to their performance. Fig. \ref{fig:switching-diagram} illustrates the proposed online switching control algorithm. 
We prove the online switching algorithm can maintain the closed-loop stability guarantees and demonstrate significant performance improvement compared to using a pre-trained controller for constant inertia without online switching.

We summarize the main contribution of this paper as follows:

\begin{itemize}
    \item We identify a key controller structure, Neural-PI control, that guarantees Exp-ISS of frequency control under \revise{switching} inertia. This is, to our knowledge, the first learning-based control algorithm that guarantees stability under \emph{nonlinear} and \emph{time-varying} frequency dynamics. 
    \item We introduce an online event-triggered switching control framework for dynamical controller selection from a set of pre-trained Neural-PI controllers, leading to improved performance compared to a fixed controller \revise{for the switching system}, while maintaining the stability guarantees. 
    \item We conduct comprehensive experiments to validate the effectiveness and efficiency of our proposed online switching algorithm with Neural-PI controllers. These experiments verify the closed-loop stability and improved performance for primary frequency control under variable inertia.
\end{itemize}

\subsection{Related Work} 
\revise{Significant progress has been made in frequency control for systems with reduced and time-invariant inertia, with growing interest in approaches for time-varying inertia. This section reviews recent advancements in frequency control for both constant and time-varying inertia, summarized in Table \ref{tab:summary}. }
\subsubsection{Frequency control under time-invariant inertia} Most existing frequency control methods are designed for systems with constant inertia, including classical droop control \cite{kundur1994power, 6702462, SIMPSONPORCO20132603, 7924418}, model predictive control (MPC)~\cite{9107487,9179017} and data-driven control~\cite{ 9721402, 9960828, 9108633,9779512,yuan2024reinforcement,cui2023structured}. The most popular method for primary frequency control using synchronous generators is droop control, which is typically a linear function of the frequency deviation (possibly with deadbands and saturation) \cite{kundur1994power,6702462}. Droop control is also adopted in inverter-connected resources to provide primary frequency control, to mimic the behavior of synchronous generators \cite{SIMPSONPORCO20132603, 7924418}. However, linear controllers can be sub-optimal since the frequency dynamics are nonlinear~\cite{9179017}. Facing this challenge, MPC-based approaches 
\cite{9107487,9179017} synthesize nonlinear controllers through optimization, which can lead to computational challenges for real-time control. Considering the nonlinear nature of frequency dynamics and the requirement for fast computation, recently, reinforcement learning (RL) approaches have been proposed~\cite{9960828, 9108633}. See~\cite{9721402} for a recent review. The basic idea of RL lies in finding a policy that computes the optimal action based on observed states, aiming to maximize cumulative rewards through interaction with the environment. The key challenge with those learning-based methods is ensuring stability, which is critical for power system applications. To this end, recent studies have integrated RL with stability guarantees \cite{9779512,yuan2024reinforcement,cui2023structured}, for frequency control under constant inertia. 

\subsubsection{Frequency control under variable inertia} 
There is growing interest in frequency control under variable inertia, due to the increasing penetration of renewable generation. \cite{8619580} firstly proposed to use a switched-affine system model with the linear approximated frequency dynamics for each inertia model. Building on this model,~\cite{8973437} developed a stable, time-invariant linear controller learned from datasets of optimal time-varying LQR controllers. 
Additionally, \cite{9790800} validated the feasibility of solutions within the switched-affine system framework, leveraging the specific structure of linearized frequency dynamics. \revise{Building on this, \cite{10543148} introduces a nonlinear residual in addition to a stable linear controller to improve performance, with stability ensured through projection. }\cite{8571607} proposes a robust controller that optimizes the worst-case system performance via a $\mathcal{H}_\infty$ loop shaping controller that adapts to time-varying inertia. \cite{guo2019performance} considers the variable inertia by modeling the dynamics as a linearized stochastic swing equation, where inertia is modeled as multiplicative noise. \revise{\cite{9161355} proposes a PI-based load frequency controller that is robust to inertia change, and where stability is validated by small-signal analysis. }Despite these advancements, a common limitation persists: the models rely on a linear swing equation for modeling the frequency dynamics and therefore are unable to accurately capture the nonlinear behavior, which might compromise control performance.
There is also work on system inertia estimation for frequency control under variable inertia \cite{9296965}, employing real-time inertia estimation to determine minimum PV power reserve requirements. However, this approach does not provide stability guarantees for the switching system, and as noted in a recent survey \cite{TAN2022107362}, fast and robust inertia estimation is challenging.
\revise{
\begin{table*}[htbp]
\centering
\caption{Literature Summary on Frequency Regulation with Constant or Time-Varying Inertia.}
\label{tab:summary}
\begin{tabular}{lcccccc}
\toprule
\multirow{2}{*}{\textbf{Reference}} & \multirow{2}{*}{\shortstack{\textbf{Time-Varying}\\\textbf{Inertia}}} & \multirow{2}{*}{\shortstack{\textbf{Nonlinear}\\\textbf{Dynamics}}}& \multirow{2}{*}{\shortstack{\textbf{Nonlinear}\\\textbf{Control}}} & \multirow{2}{*}{\shortstack{\textbf{Theoretical}\\\textbf{Guarantees}}} & \multirow{2}{*}{\shortstack{\textbf{Data-}\\\textbf{Driven}}}& \multirow{2}{*}{\shortstack{\textbf{Controller}\\\textbf{Adaption}}} \\ 
 &  &  &  &  \\ \midrule
Droop Control \cite{kundur1994power, 6702462, SIMPSONPORCO20132603, 7924418} &   &  &   &\checkmark &  \\ 
MPC \cite{9107487,9179017} &   &  & \checkmark &  \\ 
RL \cite{ 9721402,9108633,9779512} &   &\checkmark  &\checkmark   & &\checkmark \\ 
Stable RL \cite{9960828,yuan2024reinforcement,cui2023structured} &   &\checkmark  &\checkmark  &\checkmark  & \checkmark \\ 
Small Signal \cite{8619580,8973437,9790800,10543148} & \checkmark  &  & (\checkmark)~\cite{10543148}  & \checkmark&\checkmark \\ 
Robust Control \cite{9161355,8571607,guo2019performance} & \checkmark  &  &  & \checkmark \\ 
The present paper  &  \checkmark & \checkmark & \checkmark & \checkmark &\checkmark & \checkmark\\ 
\bottomrule
\end{tabular}
\end{table*}}

\emph{Notation.} We use bold symbols to represent vectors. $\sip(A):=\frac{1}{2}(A+A^\top )$ denotes the symmetric part of a square matrix $A$. $\text{diag}(\bm{c})$ represents the diagonal matrix with diagonal equal to the vector $\bm{c}$. Vectors of all ones and zeros are denoted by  $\bm{1}_n, \bm{0}_n \in \real^n$, respectively. We use the superscript $*$ to denote the equilibrium points of the variables.

\section{Model and Problem Formulation}
\subsection{Power System Model}
Consider a $n$-bus power network represented as a connected graph $(\mathcal{V},\mathcal{E})$, where buses are indexed by $i,j\in \mathcal{V}:=[n]:=\{1,...,n\}$ and the connecting lines are denoted by unordered pairs $\{i,j\}\in\mathcal{E}$. State variables are phase angles $\boldsymbol{\theta}:=(\theta_i,i\in[n])\in \mathbb{R}^n$ and frequency deviations from the nominal frequency $\boldsymbol{\omega}:=(\omega_i,i\in[n]) \in \mathbb{R}^n$. 
Since the frequency dynamics only depend on the phase angle differences, we define a change of coordinates $\delta_i:=\theta_i-\frac{1}{n}\sum_{j=1}^n\theta_j,$
where $\boldsymbol{\delta}:=(\delta_i,i\in[n])\in\mathbb{R}^n$ can be understood as phase angles in the center-of-inertia coordinates. \revise{This change of coordinates is only for analysis purposes.} Denoting a bounded disturbance of power injection from the nominal set-point $p_i$ as $\Delta d_i$ (e.g., renewable and load fluctuations),  the system dynamics can be written as: 
\begin{subequations}
\label{eq:freq_dynamics}
    \begin{align}
        \dot{\delta}_i &= \omega_i-\frac{1}{n}\sum_{j=1}^n \omega_j  \label{eq:1_a},\\
        M_i\dot{\omega}_i &= p_{i}-D_i\omega_i+u_i-\sum^n_{j=1}B_{ij}\sin(\delta_i-\delta_j) + \Delta d_i, \label{eq:1_b}
    \end{align}
\end{subequations}
where $M:=\text{diag}(M_i,i\in[n])\in\mathbb{R}^{n\times n}$ are the generator inertia, $D:=\text{diag}(D_i,i\in[n])\in\mathbb{R}^{n\times n}$ are the combined frequency response coefficients from synchronous generators and frequency sensitive loads, $\bm{p}:=(p_{i},i\in[n])\in \mathbb{R}^n$ are the net power injections, $B:=[B_{ij}]\in \mathbb{R}^{n\times n}$ is the susceptance matrix with $B_{ij}=0,\forall \{i,j\}\notin\mathcal{E}$, and $\bm{u}:=(u_i,i\in[n])\in\mathbb{R}^n$ are the control actions, denoting the active power injection to regulate the frequency. Note that the time index $t$ for all the state and action variables $\delta_i(t), \omega_i(t), u_i(t)$, and the disturbance $\Delta d_i(t)$ in \eqref{eq:freq_dynamics} are omitted for brevity.

Following the recent NREL report {\cite[Fig. 13 \& Fig. 14]{nrel_inertia_report}}, the amount of inertia is piece-wise \revise{constant} because it only depends on the on/off status of synchronous generators. Thus, the frequency dynamics with time-varying inertia can be modeled as a switching system, with a predetermined set of values of equivalent inertia at each ``mode''. The evolution of the inertia in the system depends on the current mixture of conventional generators and inverter-connected resources\revise{, and the switches happen hourly }(when a large renewable generation is available, some synchronous generators will be offline). {Considering $m$ different inertia modes, the inertia} follows a piece-wise continuous switching signal $q(t): [0,\infty) \mapsto \{1, ..., m\}$ that remains unknown to the controller. 
Thus the frequency dynamics in \eqref{eq:1_b} under time-varying inertia can be written as,

\begin{align}
\label{eq:freq_dynamics_variable}
    M_{i, q(t)}\dot{\omega}_i \!=\! p_{i} \!-\! D_i\omega_i \!+\! u_i \!-\! \sum^n_{j=1}B_{ij}\sin(\delta_i-\delta_j) +\Delta d_i.
\end{align}
This dynamics model in \eqref{eq:freq_dynamics_variable} makes the following assumptions that are commonly adopted in the literature, cf.~\cite{7944568}:
\begin{itemize}
    \item Lossless lines: the line resistance is zero for all 
    $\{i,j\}\in \mathcal{E}$;
    \item Constant voltage profile: the bus voltage magnitudes for all buses are constant and equal to $1$ p.u. Reactive power flows are not considered;
    \item Bounded angle difference: the equilibrium bus phase angle difference is within $\pm \frac{\pi}{2}$ for all $\{i,j\}\in \mathcal{E}$.
\end{itemize}
Further, we assume that the bounded disturbances $\bm{\Delta d}$ do not depend on the history of states and actions. {The assumption is realistic considering disturbances arise from external, unpredictable factors like weather or sudden load changes.}

\subsection{Control System Architecture}
\begin{figure}[htbp]
    \centering
    \includegraphics[width=0.44\textwidth]{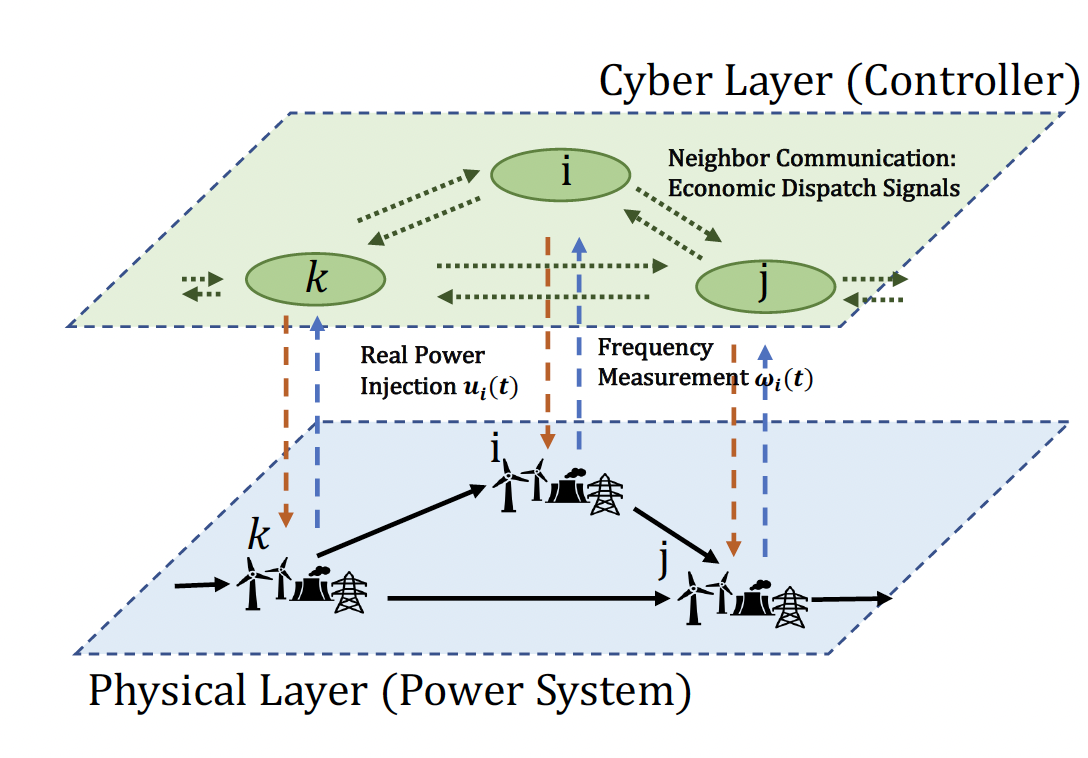}
    \caption{Diagram of the control system architecture.}
    \label{fig:control arch}
\end{figure}

We present the control system architecture in Figure \ref{fig:control arch}. \revise{The controller of inverter $i$ infers the frequency deviation $\omega_{pll,i}$ from the phase-locked-loop (PLL) block. Considering the much faster time response of PLLs on inverters (they lock around 1 AC cycle~\cite{arruda2001pll,purba2018reduced}), we assume that the measured frequency deviation $\omega_{pll,i}$ accurately approximates $\omega_i$.} To guarantee convergence to the nominal frequency and improve economic efficiency, we consider distributed communication allowing for bidirectional information exchange between neighboring buses, defined by the incidence matrix~$E$. \revise{The elements of matrix $E$ can be $+1$, $-1$, or $0$, and their values indicate the orientation of the edges relative to the buses. } The control action at bus $i$ is defined as $u_i$, which is the real power injection at bus $i$ calculated with real-time local frequency measurements $\omega_i$ and communication. 
We assume \revise{real-time communication and }timescale separation between the inverter dynamics and frequency control.

\section{Problem Formulation}
This paper focuses on frequency control with variable inertia. The switching dynamics for inertia are \emph{unknown} to the control algorithm. 
Our objective is to minimize the frequency deviation of the system while maintaining moderate control costs. 

The frequency control problem is defined as follows, 
\begin{subequations}\label{eq:dynamics_with_noise}
    \begin{align}
        \min_{\mathbf{u}}\quad& J_T= \frac{1}{T} \! \int_0^T \!\! \Big( \sum_{i=1}^n \frac{1}{2}c_i u_i^2 \!+\!\lambda(\lVert\boldsymbol{\omega}\rVert_2\!+\!\lVert\boldsymbol{\omega}\rVert_\infty) \!\Big)dt, \label{eq:objective}\\
        s.t. \quad& \dot{\mathbf{\delta}}_i = \omega_i-\frac{1}{n}\sum_{j=1}^n \omega_j, \label{opt_dyn1}\\
        &M_{i,q(t)}\dot{\omega}_i = p_{i}-D_i\omega_i+u_{i}-\nonumber
        \\ & \qquad\qquad \quad \sum^n_{j=1}B_{ij}\sin(\delta_i-\delta_j) +\Delta d_i, \label{opt_dyn2}\\
        &u_{i}=\hat{\pi}_{i} (\omega_i,s_i,\{s_j,j\in\mathcal{N}_i\})\text{ is stabilizing}\,. \label{eq:stability_constraint}
    \end{align} 
\end{subequations}

The objective function \eqref{eq:objective} encodes the control costs and a summation of 2-norm and $\infty$-norm of the frequency deviations, for reducing both the operational cost and frequency disturbance. Here, $\lambda$ is a coefficient that trades off control cost versus frequency deviation and $c_i > 0$ is the controller cost coefficient at bus $i$. 
The time-varying frequency dynamics are given in \eqref{opt_dyn1}-\eqref{opt_dyn2}. 
$\hat{\pi}_i(\omega_i,s_i,\{s_j,j\in\mathcal{N}_i\})$ is the controller at bus $i$, which requires the local frequency measurement $\omega_i$ and variables $\{s_j,j\in\mathcal{N}_i\}$ from node $i$'s neighbors $\mathcal{N}_i$. The variable $s_i\in\mathbb{R}$ includes the integral of frequency deviation and the gradient of control cost, as defined in Section \ref{sec:controller}.

To formalize the stability constraint in \eqref{eq:stability_constraint}, we employ the notion of input-to-state stability (ISS), which is commonly used in nonlinear systems with disturbances~\cite{agrachev2008input}. Specifically, we consider exponential-ISS (Exp-ISS) for fast frequency stabilization.
With $\bm{x}=\begin{bmatrix}
    \boldsymbol{\delta}-\boldsymbol{\delta}^*&\boldsymbol{\omega}-\bomega^*&k\bm{s}-k\bm{s^*}
\end{bmatrix}^\top $, {where $k$ is a control gain for the integral variable $s$ to be defined later}, we present the definition of Exp-ISS as follows. 
\begin{definition}[Exponential Input-to-State Stability (Exp-ISS)] \label{def:exp-ISS}
    A controller $\pi$ is called Exp-ISS with parameters $(\kappa,\rho,\beta)$ if, for any initial condition $\bm{x_0}\in\mathbb{R}^{3n}$ and bounded disturbance 
    $\bm{\Delta d}(t)$, i.e., $\lVert \bm{\Delta d}(t) \rVert_{L_{\infty}} =\sup_{t \ge 0}\lVert \bm{\Delta d}(t) \rVert_2$ is finite, the states satisfy
    $$\lVert \bm{x}(t)\rVert_2\leq \kappa \rho^t\lVert \bm{x_0}\rVert_2+\beta\lVert \bm{\Delta d}(t) \rVert_{L_{\infty}},$$
    for all $t\geq 0$, where $\kappa,\beta>0$, $0<\rho<1$. 
\end{definition}
\revise{Exp-ISS is a property that describes the behavior of dynamical systems in response to external inputs, which generalizes the idea of stability to systems that are not only influenced by their initial conditions but also by external disturbances. When a system is Exp-ISS, it implies that the system not only remains bounded under small external disturbances but also converges back to equilibrium exponentially fast when the input is removed. This property guarantees that the system recovers from disturbances quickly and predictably.
\begin{remark}[Robustness to measurements error, delay, or cyber attacks] \label{remark:robust}
{\rm
In practical terms, Exp-ISS guarantees the system robustness to and recovery from external disturbances, which may include unexpected load changes, or even latency, measurement errors, and certain types of cyber attacks. As long as the disturbances on the system resulting from these factors are bounded, their impact on a system with Exp-ISS guarantees
is bounded by $\beta\lVert \bm{\Delta d}(t) \rVert_{L_{\infty}}$. Furthermore,  the system will be restored exponentially fast once the disturbance is removed.
}
\oprocend
\end{remark}
}

Moreover, provided that appropriate conditions are met, the Exp-ISS guarantees for fixed subsystems can be effectively extended to a switching system that alternates between these subsystems \cite{VU2007639}.

\section{Exponentially Stable Controller Design Under All Modes} \label{sec:controller}
In this section, we design a controller structure that satisfies the Exp-ISS property in Definition \ref{def:exp-ISS} under all modes and provide a theoretical analysis. Our goal is to develop a uniform stabilizing controller for all inertia modes in the switching system that simultaneously achieves two key objectives. Firstly, restore the system to its nominal frequency, where $\bomega^* = \bm{0}_n$.
Secondly, optimize the control cost for maintaining system operation at the equilibrium, expressed as $c(\bm{u}) = \sum_{i=1}^n \frac{1}{2}c_i(u_i)^2$. 

\subsection{Controller Structure}
\revise{The steady-state optimization problem can be written as the optimal steady-state economic dispatch problem:
\begin{subequations}
    \begin{align}
    \label{eq:economic_dispath}
        \min_{\mathbf{u}^*}&\,\,\, c(\mathbf{u}^*)=\sum_{i=1}^n\frac{1}{2}c_i(u_i^*)^2\\
        \text{s.t.}&\,\,\, \sum_i (p_i+u_i^*)=0.
    \end{align}
\end{subequations}
To meet the requirement of nominal frequency restoration and steady-state optimality, Distributed Averaging-based Integral (DAI) \cite{9869334} is an established choice for frequency control.
}
\revise{Based on this idea, }we propose to use the following \emph{neural proportional-integral} (Neural-PI) controller structure, 

{
\begin{subequations}\label{eq:pi_control_nodal}
    \begin{align}
        &\hat{\pi}_{i} (\omega_i,s_i,\{s_j,j\in\mathcal{N}_i\})= \underbrace{-\pi_i(\omega_i)}_{\text{proportional term}}+ \underbrace{k {s}_i}_{\text{integral term}},\\
        &\dot{{s}_i}=-c_i^{-1}{\omega_i}-\sum_{j\in\mathcal{N}_i}(c_iks_i-c_jks_j). \label{eq:intergal_control}
    \end{align}
\end{subequations}
}

\begin{figure}[ht]
    \centering
    \includegraphics[width=0.5\textwidth]{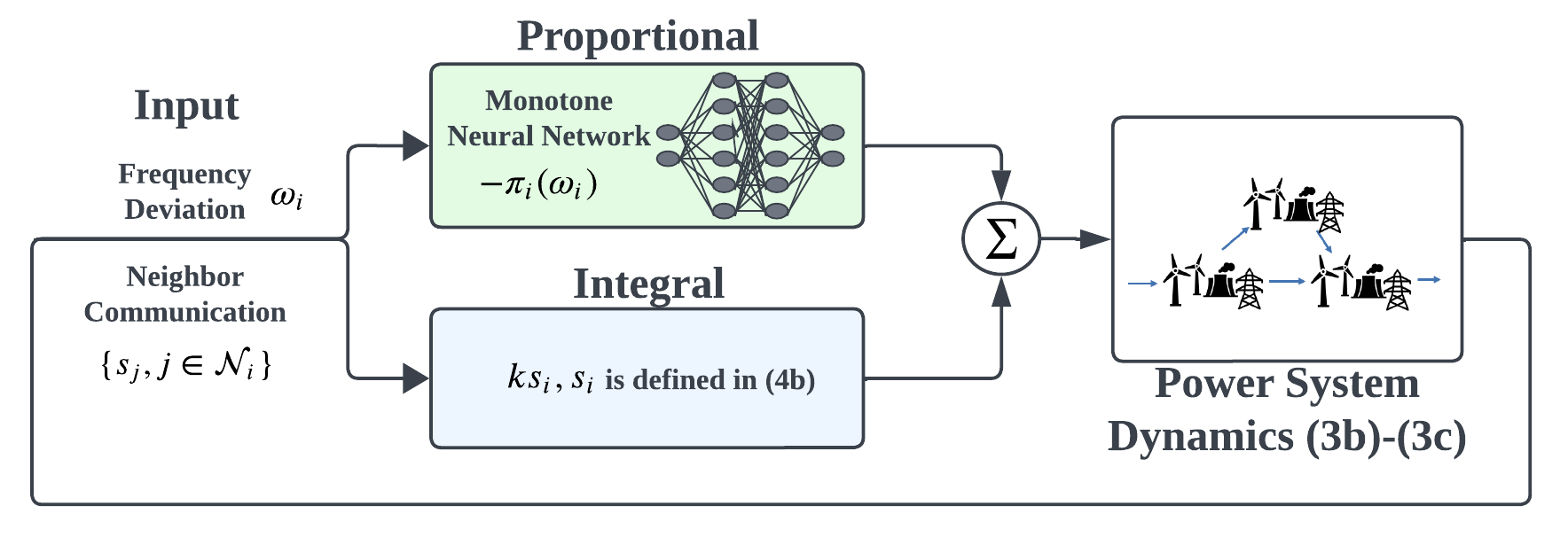}
    \caption{Diagram of the Neural-PI controller defined by \eqref{eq:pi_control_nodal}. 
    }
    \label{fig:neural-pi}
\end{figure}

We name it neural because the proportional term 
$\pi_i(\cdot)$ is a monotonically increasing function of the instantaneous frequency deviation $\omega_i$ parameterized as a {monotone neural network} with $\pi_i(0) = 0$. We name it integral because it is a linear function of $s_i$, the integral of frequency deviations and the difference in the gradients of the control cost $c(k\bm{s})$ between its neighbors. $k>0$ is a learnable control gain defined as a scalar. \revise{The linear integral control is deployed instead of a neural network to achieve exponential stability.} Figure \ref{fig:neural-pi} shows the proposed controller structure. 

Define $C$ as a diagonal matrix with diagonal entries equal $\{c_i\}_{i \in \mathcal{V}}$ and $c_i>0$ is the controller cost coefficient, we summarize the controller design \eqref{eq:pi_control_nodal} in vector form as 
\begin{subequations}\label{eq:pi_control}
    \begin{align}
        \hat{\pi}(\bomega,\bm{s})&= \underbrace{-\pi(\boldsymbol{\omega})}_{\text{proportional term}}+ \underbrace{k \bm{s}}_{\text{integral term}},\\
        \dot{\bm{s}}&=-C^{-1}\boldsymbol{\omega}-EE^\top Ck\bm{s}. \label{eq:intergal_control}
    \end{align}
\end{subequations}

\begin{remark}[Comparison to the Neural-PI controller in~\cite{cui2023structured}]
{\rm The above controller structure ~\eqref{eq:pi_control} is inspired by our earlier work~\cite{cui2023structured} for frequency control under time-invariant inertia with only asymptotic stability guarantees, where both the proportional and integral terms are parameterized as monotone neural networks. In this work, we adapt this structure to frequency control under \emph{time-varying} inertia with one major difference: only the proportional term is parameterized as a monotone neural network, and the integral term is parameterized as a linear function, in order to achieve exponential input-to-state stability guarantees (Theorem \ref{thm:iss}).
The theoretical contribution is that we extend the asymptotic stability guarantees of the Neural-PI controller to exponential ISS guarantee of each mode (Theorem \ref{thm:iss}), which is essential for establishing the stability of the switching system under different inertia.}

\oprocend
\end{remark}

\subsection{Theoretical Results}
Combining the frequency dynamics and the proposed Neural-PI controller in \eqref{eq:pi_control}, the overall closed-loop system can be modeled in vector form as follows, 
\begin{subequations}\label{eq:closed-loop}
    \begin{align}
        &\dot{\bdelta}=(I_n-\frac{1}{n}\bm{1}_n\bm{1}_n^\top )\bomega,\\
        &M_{q(t)}\dot{\bomega}=\bm{p}+\hat{\pi}(\bomega,\bm{s})-D\bomega-EB \sin(E^\top \bdelta)+\bm{\Delta d}.
    \end{align}
\end{subequations} 
Our first result characterizes the equilibrium points of the closed-loop system \eqref{eq:closed-loop} with $\bm{\Delta d}=\bm{0}_n$.

\begin{theorem}[Unique Closed-loop Equilibrium]
\label{thm:equilibrium} 
Assume $\forall \{i,j\}\!\in\! \mathcal{E}\,\, |\delta_i-\delta_j|<\frac{\pi}{2}$, the power flow equation \eqref{opt_dyn2} is feasible, and proportional term equals to zero when the frequency deviation is zero, i.e., $\pi(\bm{0}_n)=\bm{0}_n$. Then the equilibrium $(\bdelta^*,\bomega^*,k\bm{s}^*)$ of the closed-loop system \eqref{eq:closed-loop} with $\bm{\Delta d}=\bm{0}_n$ is the unique point satisfying
\begin{subequations}
    \begin{align}
        &\bomega^*=\bm{0}_n,\label{eq:equla}\\
        &EB\sin(E^\top \bdelta^*) = \bm{p}+k\bm{s^*},\label{eq:equlb}\\
        &k\bm{s^*}=\gamma C^{-1}\bm{1}_n,\label{eq:equlc}
    \end{align}\label{eq:equilibrium}
\end{subequations}
where $\gamma$ is determined by 
\begin{equation}\label{eq:lambda}
    \gamma=-\frac{\sum_{i=1}^np_i}{\sum_{i=1}^n c^{-1}_i}\,.
\end{equation}
\end{theorem}

Equation \eqref{eq:equla} indicates that the proposed controller effectively reduces frequency deviations to zero at the steady state. Given that $\bm{1}^\top _nE=0$, and upon premultiplying equation \eqref{eq:equlb}, it follows that 
$\sum_i(p_i+ks^*_i)=0$. By \eqref{eq:equlc}, the final control action, $ks^*_i$, is distributed proportionally to $\gamma c^{-1}_i$. This allocation strategy facilitates higher real power injection from lower-cost buses, thereby restoring generation balance in an economically efficient manner. Moreover, the equilibrium point remains the same regardless of the inertia mode change and switching of controllers.

For each fixed inertia mode, the following result provides an exponential ISS guarantee for the closed-loop system.

\begin{theorem}[Exp-ISS of Neural-PI Controller for Frequency Control with Time-invariant Inertia]\label{thm:iss}
Let $\pi(\bm{0}_n)=\bm{0}_n$ and $\pi_i(\omega_i)$ be monotonically increasing with respect to $\omega_i$. Consider $\mathcal{D}:=\{\bm{x}\in \real^{3n}, \forall \{i,j\}\in \mathcal{E}, |\delta_i-\delta_j|<\frac{\pi}{2}\}$. If a Neural-PI controller defined by \eqref{eq:pi_control} is deployed, then for any $\bm{x}_0\in\mathcal{D}$, the closed-loop system \eqref{eq:freq_dynamics} is exponentially input-to-state stable (Exp-ISS), i.e., there exists positive scalars $\kappa, \beta$ and $0<\rho<1$\footnote{An explicit expression of $\kappa, \rho,\beta$ is provided in \eqref{eq:detail_iss} below.} such that for all $t\geq 0$,
$$\lVert \bm{x}(t)\rVert_2\leq \kappa \rho^t \lVert \bm{x}_0\rVert_2+\beta\lVert \bm{\Delta d}(t) \rVert_{L_{\infty}}.$$
\end{theorem} 
{Theorem \ref{thm:iss} guarantees Exp-ISS for the closed-loop \eqref{eq:freq_dynamics} with the Neural-PI controller under any timer-invariant inertia, ensuring stability despite disturbances and supporting the generalization of stability to the switching system \eqref{eq:closed-loop}. We prove this using Lyapunov stability analysis by identifying a well-defined Lyapunov function that exponentially converges along the system's trajectories \eqref{eq:freq_dynamics} with bounded disturbance errors. The detailed proof is available in Appendix \ref{apdx:proof_of_thm2}.}
\revise{
\subsection{Monotone Neural Network}
The monotonically increasing function $\pi_i(\cdot)$ is constructed as a stacked ReLU function as follows.
}
\begin{corollary}\revise{(Stacked ReLU Monotone Network~\cite[Lemma 5]{9779512})}
\label{corollary:stacked_relu}
 The stacked ReLU function with $d$ hidden units constructed by 
\begin{subequations}\label{eq:relu_pos}
    \begin{align}
    &\pi^{+}(x; w^+, b^+) = {(w^+)^\top} \text{ReLU}(\mathbf{1} x + b^+)\\
    &\sum_{l=1}^{d'} w^{+}_l > 0, \forall d' = 1, ..., d\,, b^{+}_1 = 0, b^{+}_l \leq b^{+}_{l-1}, \forall l =2, ..., d
    \end{align}
\end{subequations}
is monotonically increasing for $x > 0$ and zero when $x \leq 0$. In addition, the stacked ReLU function with $d$ hidden units constructed by 
\begin{subequations}\label{eq:relu_neg}
    \begin{align}
    &\pi^{-}(x; w^{-}, b^{-}) = (w^{-})^\top \text{ReLU}(-\mathbf{1} x + b^{-})\\
    & \sum_{l=1}^{d'} w^{-}_l < 0, \forall d' = 1, ..., d\,, b^{-}_1 = 0, b^{-}_l \leq b^{-}_{l-1}, \forall l =2, ..., d
    \end{align}
\end{subequations}
is monotonically increasing for $x < 0$ and zero when $x \geq 0$.
\end{corollary}

\revise{In this way, the stacked ReLU function is a piece-wise linear function. The slope for each piece of $\pi^{+}(x; w^+, b^+)=\sum_{l=1}^{d'} w^{+}_l$, which is always positive by construction and thus satisfies the monotonic property. A similar argument holds for $\pi^{-}(x; w^{-}, b^{-})$. As a result, let $\pi(\cdot)=\pi^{+}(\cdot; w^+, b^+)+\pi^{-}(\cdot; w^{-}, b^{-})$, $\pi(x)$ is monotonically increasing. By \cite[Theorem 2]{9779512}, any continuous, Lipschitz and bounded monotonic function $r(x)$ with bounded derivatives, $r(0) = 0$, and mapping the compact set $\mathbb{X}$ to $\mathbb{R}$ can be approximated arbitrarily well by $\pi(x)$.}

\section{Online Switching Algorithm}
In this section, we introduce an online event-triggered switching algorithm with stability guarantees for frequency control under time-varying inertia. The proposed algorithm dynamically chooses from a set of pre-trained Neural-PI controllers that are optimized for different uni-inertia modes, to improve the controller performance while maintaining stability. For implementation purposes, the algorithm is presented in a discrete-time manner.

\subsection{Online Event-triggered Switching Algorithm}

\begin{figure}[htb]
    \centering
    \includegraphics[width=0.46\textwidth]{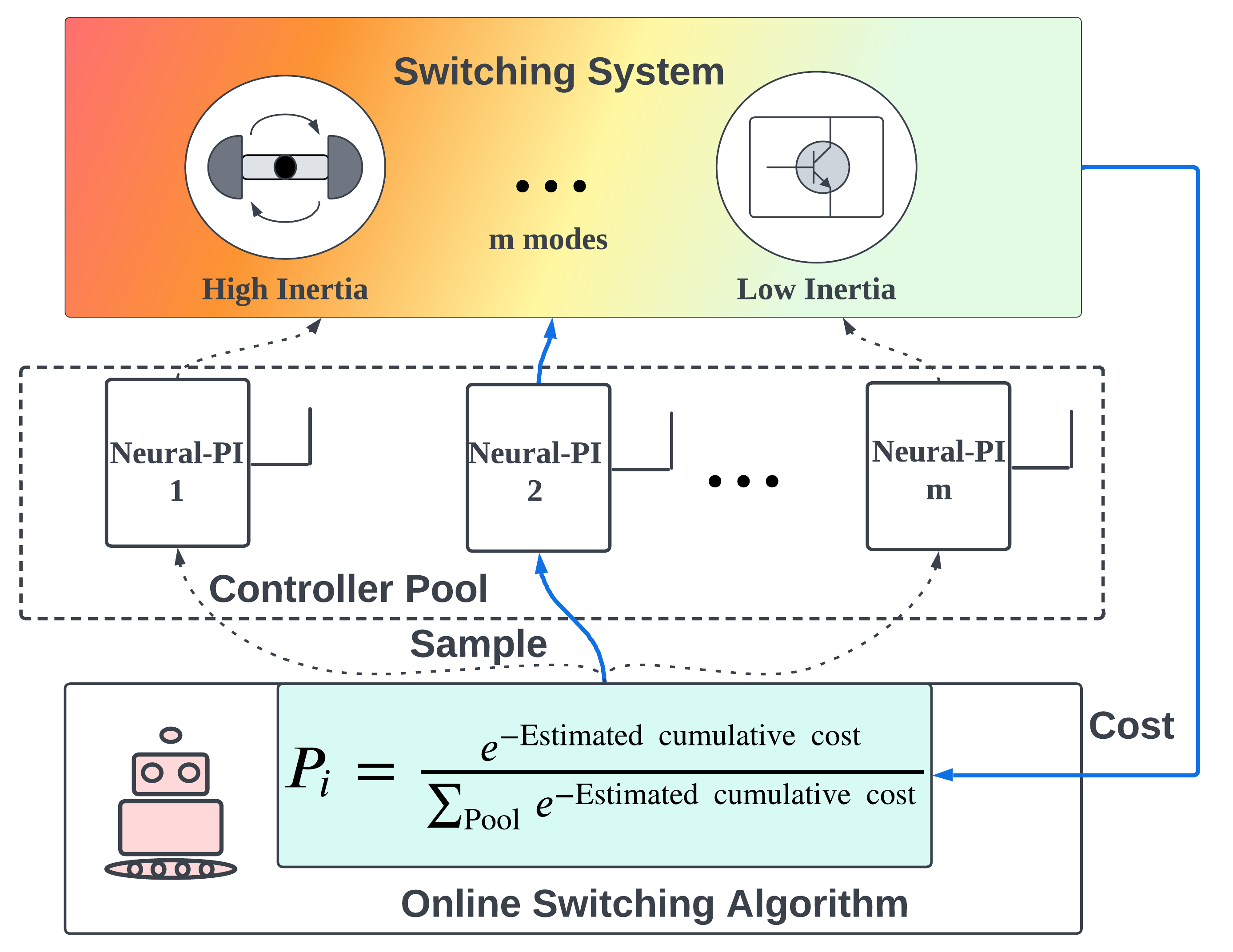}
    \caption{Online switching control for frequency control under variable inertia.}
    \label{fig:switching-alg2}
\end{figure}

Based on the previous discussion, the Neural-PI controller is capable of maintaining frequency stability under any inertia level. 
However, controllers optimized for high-inertia systems may underperform when inertia is low due to faster frequency dynamics \cite{ULBIG20147290}. 
Consequently, a uniform Neural-PI controller, even trained for all modes, compromises between different inertia modes and can result in suboptimal performance. Instead, here we propose an innovative switching algorithm to select the most appropriate controller based on the current system state from a set of Neural-PI controllers, each trained for a specific inertia level, while guaranteeing stability. The switching algorithm can improve control performance compared to base controllers trained for each specific mode. Figure \ref{fig:switching-alg2} illustrates the proposed online switching control idea. 

We now detail the online event-triggered switching algorithm in Algorithm \ref{alg:exp3}. A finite pool of candidate controllers is considered. With a slight abuse of notation, we define the index set of base Neural-PI controllers as $\mathcal{P}=\{1,\cdots,m\}$, where index $i\in\mathcal{P}$ refers to a Neural-PI controller trained in inertia mode $i$. {For theoretical purposes, the base controllers share the same integral controller gain $k$.} The proposed online switching control algorithm contains three phases: the \emph{selection phase} ($n_s$ steps), the \emph{trial phase} ($n_t$ steps), and the \emph{deployment phase}. 
The transition into the selection phase occurs once a frequency deviation exceeding $0.01$ Hz event is detected. Note that a different event-triggering threshold of frequency deviation can be chosen by the system operator according to the operation requirements.

\begin{algorithm}[t]
	\caption{Online event-triggered switching algorithm. 
 }
	\label{alg:exp3}
	\begin{algorithmic}[1]
	\Ensure Choose selection phase duration $n_s$, trial phase duration $n_t$, learning rate $\xi>0$, batch length $\tau$.
        \State{Initialize the controller selection probability $\mathrm{P}_i = 1/m$ and accumulated cost $\Tilde{G}_{i}(-1) = 0$ for  $i \in [1, 2, ..., m]$ in the controller pool; set selection flag as False and $t=t_0 =0$.}
        \While {time step $t=0,1,2,....$}
        \State{Measure frequency deviation $\bomega(t)$;}
        \If{$\lVert \bomega(t) \rVert_\infty>0.01$Hz and selection flag is False}
        \State{Set selection flag as True;}
        \EndIf
        \If{selection flag is True}
        \State{Let $t_0=t$;}
        \For {batch $j=0,1,...,\ceil{\frac{n_s}{\tau}}$ (selection phase)}
	  \State{Select controller $I_j \in \mathcal{P}$ from the selection probability $\bm{\mathrm{P}}$;}
        \State{Compute $t_{j+1} = \min(t_j+\tau,t_0+n_s)$, implement the chosen Neural-PI controller $I_j$ during batch time $[t_j, t_{j+1}]$ and calculate the batch cost,
        \begin{equation}\label{eq:batch_cost}
        \small{
        g_{(I_j)}(j)\!=\!\frac{1}{t_{j+1}\!-\!t_{j}} \sum_{t=t_j}^{t_{j+1}}\!\left (\! \sum_{i=1}^n \frac{c_i}{2}u_i(t)^2\!+\!\lambda(\lVert\boldsymbol{\omega}(t)\rVert_2 \!+ \!\lVert\boldsymbol{\omega}(t)\rVert_\infty)\right )\!;}
        \end{equation}}
        \State{Update the accumulated cost $\Tilde{G}_i(j)=\Tilde{G}_i{}(j-1)+\frac{g_{(I_j)}(j)}{\mathrm{P}_i}\mathbb{I}{(I_j=i)}$ for all $i\in \mathcal{P}$ and $\mathbb{I}$ is the indicator function;}
        \State{Update the controller selection probabilities, 
        \begin{equation}\label{eq:selection_prob}
            \mathrm{P}_i=\frac{\exp(-\xi \Tilde{G}_i(j))}{\sum_{k\in\mathcal{P}}\exp(-\xi \Tilde{G}_k(j))}, \forall i \in \mathcal{P};
        \end{equation}}
	   \EndFor
        \State{Let $t=t_0+n_s$, $\Tilde{G}_i(-1)=\Tilde{G}_i(\ceil{\frac{n_s}{\tau}})$ for all $i\in \mathcal{P}$;}
        \For{time step $t,t+1,...,t+n_t$ (trial phase)}
        \State{Commit to controller $I=\text{argmax}(\bm{\mathrm{P}})$;}
        \EndFor
        \State{Set the selection flag to False, $t=t+n_t$.}
        \Else { Commit to controller $I=\text{argmax}(\bm{\mathrm{P}})$, set $t=t+1$.}
        \EndIf
        
        \EndWhile
	\end{algorithmic}
\end{algorithm}

During the selection phase, we adopt a multi-arm bandit (MAB) framework for deciding the best controller to use from the set of pre-trained Neural-PI controllers. Each controller represents an `arm', the selection of a controller pulls an arm and yields a cost. Specifically, at each time step and if the selection phase flag is true, after observing the system state $\bomega$, a controller is selected for deciding the action, and the cost of the chosen action is revealed for updating the controller selection probability (so that controllers with lower costs will have higher probabilities to be chosen). We adopt the batched MAB in \cite{pmlr-v211-li23a} for updating the controller selection probability, since we would like to measure the controller performance over a time interval rather than a single step. Thus the algorithm proceeds in a batch manner with the batch length as $\tau$. The selected controller at batch $j$ as $I_j$ with the batched control costs and frequency deviation defined in \eqref{eq:batch_cost}, as $g_{(I_j)}(j)$. The controller selection probability $\bm{P}$ is updated at the end of each batch using the historical accumulated cost $\Tilde{G}_i(j)$ following the exponential weight formula in \eqref{eq:selection_prob}. After the selection phase, the most probable controller for the current mode, i.e., controller $I=\text{argmax}(\bm{P})$, is deployed for $n_t$ steps in the trial phase. If the frequency deviation is still greater than the pre-set threshold after the trial phase, we go back to the selection phase; otherwise, we stay in the deployment phase and commit to controller $I$ until the next triggering event with a large enough frequency deviation. \revise{For each selection, the complexity of the online switching algorithm is $\mathcal{O}(mn_s)$, which enables efficient real-time application.}

\subsection{Stability Guarantees for the Switching System}
With the online event-triggered switching algorithm in Algorithm \ref{alg:exp3}, we now proceed to provide stability guarantees for the switching system. Let $N_q(T,t)$ be the number of mode switches in the interval $[t,T)$. The switching signal $q(t)$ has an average dwell-time $\tau_a$ if there exists $N_o,\tau_a>0$ such that
$N_q(T,t)\leq N_o+\frac{T-t}{\tau_a}, \forall T\geq t\geq 0$. The following result states that, if inertia switches sufficiently slow (with a sufficiently large $\tau_a$), as compared to the time scale of the control, the switching system with both inertia switching and controller switching is still guaranteed to be Exp-ISS.

\begin{theorem}[Exp-ISS for the switching system]\label{thm:switching} 

Let $\pi(\bm{0}_n)=\bm{0}_n$ and $\pi_i(\omega_i)$ be monotonically increasing with respect to $\omega_i$. Consider a finite number of inertia modes $\{1,\cdots,m\}$, with each candidate controller in the pool $\mathcal{P}$ being a Neural-PI controller as defined by \eqref{eq:pi_control} deployed over $\mathcal{D}:=\{\bm{x}\in \real^{3n}, \forall \{i,j\}\in \mathcal{E}, |\delta_i-\delta_j|<\frac{\pi}{2}\}$. There exist constants $\tau_a^*, \kappa^* > 0$, $\rho^* \in (0,1)$, and $\beta^* > 0$ such that, if the average dwell time $\tau_a > \tau_a^*$, then with the online event-triggered switching Algorithm \ref{alg:exp3}, 

the switching system satisfies
\begin{equation}\label{eq:switching-exp-iss}
    \lVert \bm{x}(t)\rVert\leq \kappa^* {\rho^*}^t\lVert\bm{x_0}\rVert_2+\beta^*\lVert \bm{\Delta d}(t) \rVert_{L_{\infty}} .
\end{equation}
\end{theorem}

We provide the proof sketch for Theorem \ref{thm:switching}. Consider first the case of inertia changes only. In this case, Theorem \ref{thm:switching} is a direct consequence of \cite[Theorem 3.1]{VU2007639}.  When switching of controllers is considered, the Lyapunov function \eqref{eq:Lyapunov} is invariant to controller changes. Thus the Lyapunov function for the current controller is a common Lyapunov function for all base Neural-PI controllers, and switching of controllers retains Exp-ISS. 

Note that the average dwell time $\tau_a$ is sufficiently large in frequency control. A long average dwell time $\tau_a$ of inertia implies slow inertia switching. In scenarios where synchronous generators provide inertia, such switches occur on an hourly basis, whereas control actions are executed in second and sub-second scale~\cite{nrel_inertia_report}. Therefore, the average dwell time $\tau_a$ is sufficiently large for the controllers, and Exp-ISS for the system with switching inertia is preserved with rates $\kappa^*$, $\rho^*$ and $\beta^*$. Theorem \ref{thm:switching} also generalizes the stability guarantees of the batched MAB algorithm~\cite{pmlr-v211-li23a} from an unknown time-invariant system to unknown time-varying systems. When the switching of inertia is sufficiently slow, the online event-triggered control algorithm with the controller pool composed of Neural-PI controllers preserves the exponential ISS guarantees of the base Neural-PI controllers while enhancing the performance.

\section{Experiments}
This section first introduces the experiment setup and model training details. Then, we evaluate the performance of the base Neural-PI controllers and the proposed online switching algorithm. Finally, we study the impact of hyperparameters in the online switching algorithm through sensitivity analysis. 

\subsection{Experiment Setup}
We evaluate the performance of different controllers using the IEEE New England 10-machine 39-bus (NE39) network~\cite{4113518}. 
There are three inertia modes, where the inertia constants $M$ are set at $30\%, 100\%$, and $500\%$ of the standard values \cite{4113518}. These correspond to a low inertia scenario with prevalent renewable generation (denoted as $0.3$), a standard scenario (denoted as $1.0$), and a scenario dominated by synchronous generators (denoted as $5.0$), respectively. Three base Neural-PI controllers are trained under specific inertia levels and denoted as `Neural-PI-0.3', 'Neural-PI-1', and 'Neural-PI-5'. The Neural-PI controller structure is defined in \eqref{eq:pi_control}, where the proportional term $\pi(\cdot)$ is parameterized as a monotone neural network \revise{defined in Corollary \ref{corollary:stacked_relu}} with 1 hidden layer and 20 hidden units. {To comply with the operational constraints, we threshold our control policy with action bounds, i.e. $[\hat{\pi}_{i} (\omega_i,\{s_j,j\in\mathcal{N}_i\})]_{\underline{u}_i}^{\bar{u}_i}$, where $[\cdot]_{\underline{u}_i}^{\bar{u}_i}$ represents a projection onto $[\underline{u}_i,\bar{u}_i]$. These constraints are not considered in our theoretical analysis.} 

\revise{
\subsubsection{Training Algorithm} To optimize the Neural-PI controller for each given inertia model, we adopt the Physics-informed Reinforcement Learning with RNN structure from \cite{9779512}, where RNN is a class of neural networks designed for modeling temporal sequences. Given that all states are time-coupled, we integrate the state transition dynamics \eqref{opt_dyn1}-\eqref{opt_dyn2} of the power system to the RNN framework following the training algorithm in \cite{cui2023structured} to train the base Neural-PI policies.
    \begin{figure}
        \centering
        \includegraphics[width=\linewidth]{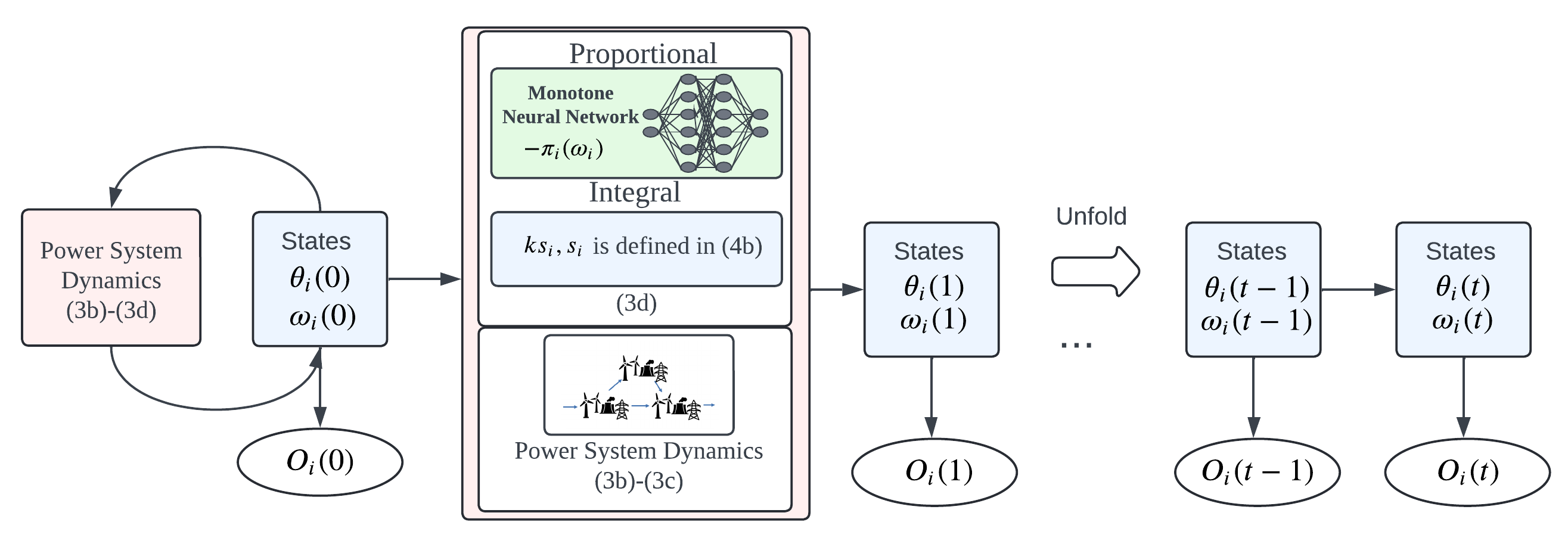}
        \caption{Structure of RNN for policy optimization.}
        \label{fig:rnn-diagram}
    \end{figure}}
    
    \revise{
    The operation of RNN is shown by the left side of Figure~\ref{fig:rnn-diagram}. The cell unit of RNN will remember its current state at the initial stage and pass it to the Neural-PI controller, where the controller gives the corresponding control action. The integrated power system dynamics give the output, which is the state at time~$1$. Then the output is fed as input to the next stage. Unfolding this process gives Figure \ref{fig:rnn-diagram}. The output at stage $t$ includes the control action $u_i(t)$ and the frequency measurement $\omega_i(t)$ for all buses. 
}
For each training trajectory, a random net-load disturbance 
is generated at a random time step. The total trajectory length is 3s with step size $\Delta t = 0.01s$ ($T=300$). To obtain the trained controllers, we run $300$ episodes, each episode containing $300$ trajectories. Parameters of the controllers are updated via gradient descent to minimize the following loss function,

\begin{equation}
    L_{\hat{\pi}} = \frac{1}{T}\sum_{t=1}^{T} \Big( \sum_{i=1}^n \frac{1}{2}c_i u_i^2(t)+\lambda(\lVert\boldsymbol{\omega}(t)\rVert_2 + \lVert\boldsymbol{\omega}(t)\rVert_\infty)\Big ).
\end{equation}
with initial learning rate $0.05$, decaying every 50 episodes with a base of $0.7$. \revise{Adam algorithm \cite{KingBa15}
is adopted as the optimizer.} 
We use the TensorFlow 2.0 framework to build the learning environment and run the training process with a single Nvidia 1080 Ti GPU with 11 GB memory. {Our code is available \href{https://github.com/JieFeng-cse/Online-Event-Triggered-Switching-for-Frequency-Control}{online}.\footnote{https://github.com/JieFeng-cse/Online-Event-Triggered-Switching-for-Frequency-Control} \revise{Training for each base Neural-PI controller takes 429.9 seconds (roughly 7 minutes), each inference of the trained Neural-PI controller takes an average of $1.29$ ms, which is sufficient for real-time implementation.}}

\revise{\subsubsection{Baselines} Besides base Neural-PI controllers trained with constant inertia, we test the proposed online switching algorithm against the following baseline algorithms.
\begin{itemize}
    \item Linear-Droop: Standard linear droop controller \cite{kundur1994power}, where the linear coefficients are learnable.
    \item Linear-PI: PI controller with linear proportional and integral control \cite{8556089,9161355}, both proportional and integral coefficients are learnable.
    \item Lyapunov-NN: Neural network proportional control policy with monotone structure design \cite{9779512}.
    \item NN-PI: A standard neural network proportional controller with linear integral control. Standard fully connected neural networks are widely used as controllers in various RL-based frequency regulation studies \cite{9721402}.
\end{itemize}
All the parameters of the baseline methods are trained with the RNN framework using the same settings as our method. Instead of using a constant inertia mode to train the controllers as the base Neural-PI controllers, all three modes are explored by the baseline models at the same time.}
With trained controllers, we first test the base Neural-PI controllers in different constant inertia modes to evaluate their transient performances in different inertia levels (Table \ref{table:base}), where each test trajectory has a 3s duration and a random net-load change at the start. We then demonstrate the efficiency of our online switching algorithm \revise{against all baseline methods} with 20s trajectories, where mode transitions occur every five seconds, which is a challenging scenario for experimental purposes (Table \ref{table:switching}). Additionally, random net-load disturbances are introduced at the $0.1$-second and $7.0$-second. 
The initial inertia is set at 0.3, 1.0, and 5.0 with respective probabilities of 10\%, 45\%, and 45\%. Inertia transitions include increasing, decreasing, or staying the same, governed by a Markov process. If the current inertia level is $0.3$, the inertia level will stay at $0.3$ or increase to $1.0$ with 50\% probability each. If the inertia level is at $1.0$, there will be a probability of 30\%, 40\%, and 30\% to decrease, stay, and increase the inertia level. If the current inertia is $5$, it follows 50\% probability to switch to $1.0$ and 50\% to stay the same. These probabilities, predetermined for experimental purposes, are \emph{unknown} to the online control algorithm. We set $\tau=5$, $\xi=5e^{-{3}}$, $n_s=50$, $n_t=300$ in Algorithm \ref{alg:exp3}.

\subsection{Performance of the Online Switching Control}
\subsubsection{Evaluation Metrics} We consider the transient period as the 3-second time interval after a disturbance and define the transient performance as the cumulative cost during this period. 
Three metrics are used to evaluate the controller performance:
\begin{itemize}
    \item \emph{Frequency deviation}: is defined as $\frac{1}{T}\sum_{t=1}^{T} \lambda(\lVert\boldsymbol{\omega}(t)\rVert_2 + \lVert\boldsymbol{\omega}(t)\rVert_\infty)$ denoting the average frequency deviation during the transient period $T = 300$;
    \item \emph{Control cost}: is defined as $\frac{1}{T}\sum_{t=1}^{T}c(\bm{u}(t))$ measuring the average control cost during the transient period $T = 300$; 
    \item \emph{Total cost}: sum of frequency deviation and control cost.
\end{itemize}
\subsubsection{Statistical Results}We first present the performance of the base Neural-PI controllers at different fixed inertia modes to illustrate the sub-optimality when a controller is trained under one inertia level but then used for a different inertia. Table \ref{table:base} summarizes the average total cost along 100 test trajectories.
\begin{table}[h]
\centering
\caption{Total control costs of different base Neural-PI controllers under different inertia modes{, with the best performance in each mode highlighted in bold.}}
 \label{table:base}
 \begin{tabular}{lcccccc}
    \toprule
    & \multicolumn{2}{c}{Inertia 0.3}  & \multicolumn{2}{c}{Inertia 1.0}  & \multicolumn{2}{c}{Inertia 5.0}\\
    \cmidrule(r){2-7}
    Method     & Mean     & Std & Mean & Std& Mean & Std \\
    \midrule
    Neural-PI-0.3  &{\textbf{22.50}}  & {8.25} &{22.74}&{8.37} &{23.81} &{8.35} \\
    Neural-PI-1 &335.07 & 2.84& \textbf{11.23}& 2.40& 11.24 & 2.34 \\
    Neural-PI-5  & 488.07 &3.03&13.10&4.47 & \textbf{10.52}&2.06\\
    \bottomrule
\end{tabular}
\end{table}

Table \ref{table:base} shows that base controllers optimized for specific inertia levels outperform others, indicating that proper switching of controllers can improve performance. 
Notably, at a low inertia level, controllers Neural-PI-1 and Neural-PI-5 result in much higher costs because of large control actions and induced frequency oscillations. As inertia increases, frequency dynamics get slower and allow larger control actions to stabilize the system. As a result, Neural-PI-1 and Neural-PI-5 outperform Neural-PI-0.3 under higher inertia levels.

\begin{table}[t]
\centering
\caption{Transient performance of the known switching, online switching, base Neural-PI controllers\revise{, and other baselines}. Known Switching defines the offline optimum assuming that the inertia switches are known to the controller, and the corresponding controller is selected once a mode change happens. {Performance of our algorithm is highlighted in bold.}}
 \label{table:switching}
 \begin{tabular}{lcccccc}
    \toprule
    & \multicolumn{2}{c}{Total Cost}& \multicolumn{2}{c}{Freq Deviation}  & \multicolumn{2}{c}{Control Cost}  \\
    \cmidrule(r){2-7}
    Method     & Mean     & Std & Mean & Std& Mean & Std \\
    \midrule
    Known Switching &  14.52&  4.46& 13.74 & 4.41 & 0.78& 0.31 \\
    \textbf{Online Switching}   &  \textbf{17.34}& \textbf{7.28} & \textbf{16.62} & \textbf{7.22} &\textbf{0.72} & \textbf{0.22}\\
    Neural-PI-0.3 &  21.91& 7.11 & 21.58 & 6.97 &0.32 & 0.16\\
   Neural-PI-1&  48.57& 74.05 & 46.50 & 71.51 &2.07& 2.55\\
   Neural-PI-5&  56.18& 101.91 & 54.13 & 99.36 &2.04 & 2.57\\
   Linear-PI & 29.10 & 8.63& 28.81 &  8.52&0.29 & 0.13\\
   NN-PI & 245.4 & 520.2 &245.0 &519.7 &0.35 &0.49 \\
   Lyapunov-NN & 22.52 & 4.62 & 22.11 & 4.49 &0.41&0.14\\
   Linear-Droop & 38.35&11.33&38.10&11.20&0.25&0.13\\
    \bottomrule
\end{tabular}
\end{table}

\begin{figure}[h]
        \centering
        \includegraphics[width=\linewidth]{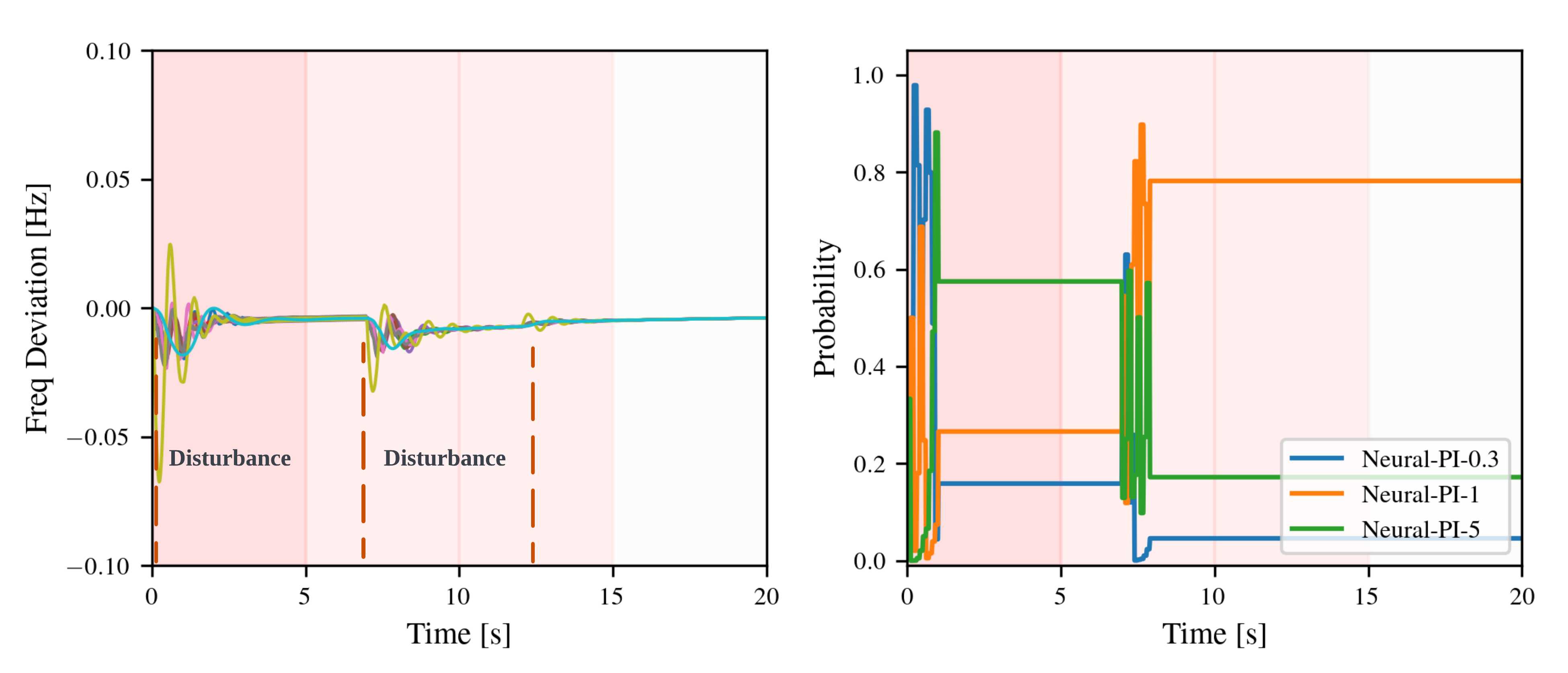}
        \caption{High-order simulation with ANDES. Control trajectory of all controlled buses with the proposed online event-triggered switching control algorithm and the evolution of controller selection probabilities.  The background colors represent the inertia levels, where the inertia switching sequence is $\{5.0,1.0,1.0,0.3\}$.}
        \label{fig:andes_all_bus}
    \end{figure}

We now evaluate the performance of \emph{known switching}, the \emph{proposed online switching control} in Algorithm~\ref{alg:exp3}, \emph{base Neural-PI controllers}, and other baselines for frequency control under variable inertia. Known switching refers to the ideal scenario where, as soon as the inertia mode changes, the corresponding controller is deployed instantly. However, real-time mode detection poses significant challenges, as highlighted by \cite{TAN2022107362}, and is thus considered an ideal performance benchmark. This experiment uses a 20-second trajectory with mode transitions occurring every five seconds. Results from 100 test trajectories are summarized in Table \ref{table:switching}. \revise{The Lyapunov-NN achieves the best performance among the four baselines except Neural-PI based methods. However, without integral control, this controller may fail to restore frequency to its nominal value and achieve steady-state optimality, and yield a sub-optimal performance. While Linear-PI can achieve similar stability guarantees, the neural network design of Neural-PI structure enables our approach to have more flexibility, leading to an improved performance. Without any stability guarantees, a general NN-PI controller can lead to frequency oscillations and a large frequency cost.}
Our online event-triggered switching control achieves the significant improvement compared to the Neural-PI baselines without switching \revise{and outperforms all other baselines}.
\revise{
\subsubsection{High-Order Power System Simulation}
To illustrate how the proposed online switching algorithm works in practice, we test the trained models and the online event-triggered algorithm with $6^{th}$ order generator model as well as dynamic models for inverter-connected resources \cite{207380,7332987}. ANDES (an open-source package for high-order power system dynamic simulation) \cite{cui2020hybrid} is utilized to simulate the dynamic response from the Western Electricity Coordinating Council (WECC) generic models \cite{farantatos2018model} for solar PV generation as the renewable resources, and $6^{th}$-order generator model with turbine-governing systems. Parameters for the PV and the voltage control follow the default values in ANDES. A test trajectory is provided in Figure \ref{fig:andes_all_bus}, illustrating controlled frequency deviation and the evolution of controller selection probability distribution $\bm{P}$ across an inertia switching sequence $\{5.0,1.0,1.0,0.3\}$. Initially, a disturbance triggers the selection phase at around 0.1 seconds. Neural-PI-5 outperforms other controllers in the controller pool, thus it is selected after the selection and trial phases, and is deployed until a new disturbance at 7 seconds prompts another selection phase, where Neural-PI-1 is chosen. With frequency disturbance restored, no further switching is triggered even if mode changes happen or another small disturbance happens at 12 seconds. As demonstrated in this high-order simulation, our method effectively restores frequency deviation after disturbances, thanks to its robustness to errors, cf. Remark \ref{remark:robust}, and correctly identifies the appropriate controller online.}

\subsection{Sensitivity Analysis}
Last, we examine the impact of hyperparameters $\xi$ (learning rate) and $\tau$ (batch length) in the proposed online switching algorithm in Algorithm~\ref{alg:exp3}. Results are presented in Table \ref{table:sensitivity}. The learning rate, $\xi$, modulates the algorithm's responsiveness to cost changes, set initially at $5e^{-4}$ based on the cost scale in our experiments. A lower $\xi$ results in a more gradual evolution of the probability distribution $\bm{P}$, avoiding rapid probability change but potentially leading to prolonged use of sub-optimal controllers. 
On the other hand, a higher $\xi$ accelerates convergence towards a particular controller, while risking being greedy and committing to a sub-optimal choice too early. 
The batch length $\tau$ also influences the balance between exploration and exploitation. 
For our experiments, with a selection phase of $n_s=50$, $\tau$ must be less than 10 to allow sufficient trials of different controllers. At $\tau=10$, the agent has merely five opportunities to evaluate different controllers, increasing the risk of prematurely converging to a sub-optimal choice. 
A $\tau$ value of 1 leads to controller update and switching at every step, potentially leading to more oscillations and short-sighted controller performance evaluation. Therefore, properly choosing hyperparameters is essential for optimizing performance.

\begin{table}[t]
\centering
\caption{Transient performance of the online switching algorithm with different hyperparameters. {Results of the selected hyperparameters are highlighted in bold.}}
 \label{table:sensitivity}
 \begin{tabular}{lcccccc}
    \toprule
    & \multicolumn{2}{c}{Total Cost}& \multicolumn{2}{c}{Freq Deviation}  & \multicolumn{2}{c}{Control Cost}  \\
    \cmidrule(r){2-7}
    Parameter     & Mean     & Std & Mean & Std& Mean & Std \\
    \midrule
    $\xi=5e^{-4},\tau=5$ &  17.92&  9.16& 17.20 & 9.03 & 0.73& 0.28 \\
     $\bm{\xi=5e^{-3},\tau=5}$   &  \textbf{17.34}& \textbf{7.28} & \textbf{16.62} & \textbf{7.22} &\textbf{0.72} & \textbf{0.22}\\
    $\xi=3e^{-2},\tau=5$  &  18.10& 8.60 & 17.39 & 8.56 &0.71 & 0.24\\
    $\xi=5e^{-3},\tau=1$&  18.28& 9.36 & 17.57 & 9.26 &0.71 & 0.29\\
    $\xi=5e^{-3},\tau=10$&  20.69& 12.30 & 19.94 & 12.16 &0.74 & 0.32\\
    \bottomrule
\end{tabular}
\end{table}

\section{Conclusions}
In this work, we have considered the problem of primary frequency control under 
time-varying system inertia. To address it, we have modeled it as a switching system, where the frequency dynamics under each mode are described by the nonlinear swing equation, and different modes represent different system inertia levels determined by the ratios of inverter-connected resources and synchronous generators. We proposed an online event-triggered switching control scheme, to dynamically select a controller online from a set of pre-trained Neural-PI controllers under specific inertia levels. Our design leverages the Exp-ISS properties of the base Neural-PI controller, to establish the Exp-ISS guarantee of the switching system. The efficacy of the proposed approach is demonstrated on the IEEE 39-bus system, where it achieves a notable reduction in average cost by approximately 20.9\% compared to the best base Neural-PI controllers in the controller pool. \revise{Additionally, in high-order power system simulations, our algorithm successfully identifies the appropriate controller online and restores frequency deviations.}
In terms of future directions, we are interested in incorporating line losses and inverter dynamics into the stability analysis, \revise{along with simultaneous consideration of voltage control}. \revise{We also aim to extend our framework to regional frequency control.} \revise{Another challenge is to specifically incorporate cyber-security and latency aspects in the proposed framework.}

\section*{Acknowledgments}
The authors would like to thank Dr. Patricia Hidalgo-Gonzalez at UC San Diego and Dr. Yingying Li at the University of Illinois Urbana-Champaign for helpful discussions.

\bibliographystyle{IEEEtran}
\bibliography{reference}
\section*{Appendices}
\addcontentsline{toc}{section}{Appendices}
\renewcommand{\thesubsection}{\Alph{subsection}}
\input{appendix}

\end{document}

%% file: appendix.tex
\subsection{Proof of Theorem \ref{thm:equilibrium}}
\begin{proof}
At steady state, the dynamics \eqref{eq:closed-loop} yields 
\begin{subequations}
\begin{align}
    &\bomega^* = \bm{1}_n\omega^*,\\
    &\bm{0}_n= \bm{p}+\bm{u}(\bomega^*,\bm{s}^*)-D\bomega^*-EB\sin(E^\top \bdelta^*),\label{eq:eq_b}\\
    &C^{-1}\bomega^* = -EE^\top Ck\bm{s^*}, \label{eq:eq_c}
\end{align}   
\end{subequations}
where $\omega^*$ is a scalar. Premultiplying \eqref{eq:eq_c} by $\bm{1}_n^\top $ yields 
$$\bm{1}_n^\top C^{-1}\bm{1}_n\omega^*=-\bm{1}_n^\top EE^\top Ck\bm{s^*}=0,$$
which implies that $\omega^*=0$. As a result, the right-hand side of \eqref{eq:eq_c} also equals the zero vector. Following \cite[Lemma 5]{cui2023structured},  $EE^\top  Ck\bm{s}=\bm{0}_n$ if and only if $Ck\bm{s}\in \text{range}(\bm{1}_n)$, which indicates that $Ck\bm{s^*}=\gamma \bm{1}_n$, implying that $k\bm{s^*}=\gamma c^{-1}\bm{1}_n$. In this case, $\bm{u}^*=k\bm{s^*}$ is the unique minimizer of the optimal steady-state economic dispatch problem following \cite{9869334}. Applying these results to \eqref{eq:eq_b} yields
$$EB\sin(E^\top \boldsymbol{\delta^*})=\bm{p}+k\bm{s^*}=\bm{p}+\gamma C^{-1}\bm{1}_n.$$
By premultiplying the above equation with $\bm{1}_n^\top $ and the fact that $\bm{1}_n^\top EB\sin(E^\top \boldsymbol{\delta}^*)=0$, we can reach to \eqref{eq:lambda}. This result is similar to the analysis in \cite[Theorem 1]{9869334}. Given that the inertia does not show up in the equilibrium analysis, the switching system has a unique equilibrium point. 
\end{proof}
\subsection{Proof of Theorem \ref{thm:iss}}  \label{apdx:proof_of_thm2}
We use the Lyapunov stability theory. The proof is structured as follows. We first present the Lyapunov function for the closed-loop system, which can be bounded by the norm of $\bm{x}$ on both sides. We then bound its Lie derivative under disturbances. Following the application of the Comparison Lemma~\cite[Adopted from Lemma 3.4]{khalil2002nonlinear},
we arrive at Exp-ISS.

\textit{Lyapunov Function:} Define the function $R(s)$ as
\begin{equation}
    R(s):=\sum_{i=1}^n c_i\int^{s_i}_0 kzdz=\sum_{i=1}^n \frac{c_ik}{2} s_i^2.
\end{equation}
Then, the Bregman distance is defined as $B^{\mathcal{V}}(\bm{s},\bm{s}^*)$, i.e.,
\begin{equation}
    B^{\mathcal{V}}(\bm{s},\bm{s}^*):=R(\bm{s})-R(\bm{s}^*)-\nabla R(\bm{s}^*)^\top (\bm{s}-\bm{s}^*)
\end{equation}
Notably, the Bregman distance here simplified as $B^\mathcal{V}(\bm{s},\bm{s^*})=\frac{k}{2}\sum_i c_i(s_i-s_i^*)^2$.
We consider the following Lyapunov function candidate for a fixed inertia $M$:
\begin{align}\label{eq:Lyapunov}
    &V(\boldsymbol{\delta},\boldsymbol{\omega},\bm{s})=\frac{1}{2}\sum_{i=1}^n M_i(\omega_i)^2+W_p(\boldsymbol{\delta}) \\
    &+\epsilon_1 W_c(\boldsymbol{\delta},\boldsymbol{\omega})+B^\mathcal{V}(\bm{s},\bm{s}^*) - \epsilon_2 (\bm{s}-\bm{s}^*)^\top \bm{1}_n\bm{1}_n^\top M\bomega, \nonumber
\end{align}
where 
\begin{subequations}
    \begin{align*}
        W_p(\boldsymbol{\delta}):&=-\frac{1}{2}\sum_{i=1}^n\sum_{j=1}^nB_{ij}(\cos(\delta_{ij})-\cos(\delta^*_{ij}))
        \\
        &-\sum_{i=1}^n\sum_{j=1}^n B_{ij}\sin(\delta_{ij}^*)(\delta_i-\delta_i^*),
    \end{align*}
\end{subequations}
$$W_c(\boldsymbol{\delta},\boldsymbol{\omega}):=\sum_{i=1}^n\sum_{j=1}^n B_{ij}(\sin(\delta_{ij})-\sin(\delta_{ij}^*))c_iM_i(\omega_i-\omega^*),$$
and $\delta_{ij}:=\delta_i-\delta_j$. Here, $\epsilon_1,\epsilon_2>0$ are tunable parameters. The last cross-term in~\eqref{eq:Lyapunov} is inspired by \cite{WEITENBERG2018103}.
\begin{lemma}[Bounds on Lyapunov Function]\label{lm1}
    For all $(\boldsymbol{\delta},\boldsymbol{\omega},\bm{s})\in \mathcal{D}$, there exist $a_1, a_2>0$\footnote{Explicit expressions for $a_1,a_2>0$ are given in \eqref{eq:constant_a1a2}.}, such that the Lyapunov function $V(\boldsymbol{\delta},\boldsymbol{\omega},\bm{s})$ is bounded by the following inequalities
    $$V(\boldsymbol{\delta},\boldsymbol{\omega},\bm{s})\geq a_1(\lVert \boldsymbol{\delta}-\boldsymbol{\delta}^*\rVert_2^2+\lVert \boldsymbol{\omega}-\bomega^*\rVert_2^2 + \lVert k\bm{s}-k\bm{s}^*\rVert_2^2),$$
    $$V(\boldsymbol{\delta},\boldsymbol{\omega},\bm{s})\leq a_2(\lVert \boldsymbol{\delta}-\boldsymbol{\delta}^*\rVert_2^2+\lVert \boldsymbol{\omega}-\bomega^*\rVert_2^2+ \lVert k\bm{s}-k\bm{s}^*\rVert_2^2)).$$
\end{lemma}
\begin{proof}[Proof of Lemma \ref{lm1}]
The proof is similar to the proof of \cite[Lemma 1]{WEITENBERG2018103}. We will bound $V(\bdelta,\bomega,\bm{s})$ term-by-term. First, following the Rayleigh-Ritz theorem, the kinetic energy term $\frac{1}{2}\sum_{i=1}^n M_i(\omega_i-\omega^*)^2$ is bounded with lower bound $\frac{1}{2}\lambda_{\min}(M)\lVert \boldsymbol{\omega}-\bomega^*\rVert^2_2$ and upper bound $\frac{1}{2}\lambda_{\max}(M)\lVert \boldsymbol{\omega}-\bomega^*\rVert^2_2$. Following \cite[Lemma 4]{WEITENBERG2018103}, $W_p(\bdelta)$ can be bounded by ${\small \frac{\eta_1}{2}\lVert \boldsymbol{\delta}-\boldsymbol{\delta}^*\rVert_2^2\leq W_p(\boldsymbol{\delta})\leq \frac{\eta_2}{2}\lVert \boldsymbol{\delta}-\boldsymbol{\delta}^*\rVert_2^2}$ with some positive $\eta_1,\eta_2$. Define $C_{\max}=\max({C})$, the absolute value of the cross-term $W_c(\bdelta)$ is bounded as follows,
$$|W_c(\boldsymbol{\delta})|\leq \frac{1}{2}(\eta_2^2C_{\max}^2\lVert\boldsymbol{\delta}-\boldsymbol{\delta}^*\rVert_2^2+\lambda_{\max}(M)^2\lVert\boldsymbol{\omega}-\boldsymbol{\omega}^*\rVert_2^2).$$
Let $W_2=(\bm{s}-\bm{s}^*)^\top \bm{1}_n\bm{1}_n^\top M\bomega$, similarly, we have
$${\small |W_2|\leq \frac{1}{2}(\frac{n^2}{k^2}\lVert k\bm{s}-k\bm{s}^*\lVert^2_2+\lambda_{\max}(M)^2\lVert \bm{\omega}\rVert_2^2)}.$$
Because $B^\mathcal{V}(\bm{s},\bm{s^*})=\frac{1}{2k}\sum_i c_i(ks_i-ks_i^*)^2$, it can be bounded by $\frac{C_{\min}}{2k}\lVert k\bm{s}-k\bm{s}^*\rVert_2^2\leq B^\mathcal{V}(\bm{s},\bm{s^*})\leq \frac{C_{\max}}{2k}\lVert k\bm{s}-k\bm{s}^*\rVert_2^2$, with $C_{\min}=\min(C)$.
Combining the inequalities, we can bound the entire Lyapunov function with
{\small
\begin{subequations}\label{eq:constant_a1a2}
    \begin{flalign}
    a_1:=&\frac{1}{2}\min (\lambda_{\min}(M)-(\epsilon_1+\epsilon_2) \lambda_{\max}(M)^2, \nonumber\\
    &\eta_1-\epsilon_1\eta_2^2C_{\max}^2,
    \frac{kC_{\min}-\epsilon_2 n^2}{k^2}),\\
    a_2:=&\frac{1}{2}\max (\lambda_{\max}(M)+(\epsilon_1+\epsilon_2) \lambda_{\max}(M)^2,\nonumber\\
    &\eta_2+\epsilon_1\eta_2^2C_{\max}^2,\frac{kC_{\max}+\epsilon_2 n^2}{k^2}).
\end{flalign}
\end{subequations}
}
$\epsilon_1$ and $\epsilon_2$ are sufficiently small so that $a_1, a_2$ are positive.
\end{proof}

The next result bounds the Lie derivative of~\eqref{eq:Lyapunov}.
\begin{lemma}[Time derivative]\label{lemma:derivative}
    Given the Neural-PI controller defined by \eqref{eq:pi_control}, with $\pi(\bm{0}_n)=\bm{0}_n$ and $\pi_i(\omega_i)$ monotonically increasing with respect to $\omega_i$, consider  the Lyapunov function \eqref{eq:Lyapunov}.
    There exist $\alpha_1,\alpha_2>0$\footnote{Explicit expressions for $\alpha_1,\alpha_2>0$ are provided in \eqref{eq:alpha}.} such that, for $(\bdelta,\bomega,\bm{s})\in \mathcal{D}$,        
   $$\dot{V}(\boldsymbol{\delta},\boldsymbol{\omega},\bm{s})\leq -\alpha_1V(\boldsymbol{\delta},\boldsymbol{\omega},\bm{s})+\alpha_2\lVert \bm{\Delta d} \rVert_2 \sqrt{V(\boldsymbol{\delta},\boldsymbol{\omega},\bm{s})}.$$
\end{lemma}
\begin{proof}[Proof of Lemma \ref{lemma:derivative}]
Define $p_{e,i}(\boldsymbol{\delta}):=\sum_{j=1}^n B_{ij}\sin(\delta_{ij})$, $H(\boldsymbol{\delta})=\nabla \bm{p_e}(\boldsymbol{\delta})$. It can also be written as $H(\boldsymbol{\delta})=EB \diag(\cos(E^\top \boldsymbol{\delta}))E^\top $, where $E$ is the incidence matrix. Given that $B$ is a diagonal matrix, $H(\boldsymbol{\delta})$ is a Laplacian matrix. We start by computing the partial derivatives of $V(\delta,\omega)$ with respect to each state:
{\small\begin{align*}
    \frac{\partial V}{\partial \bdelta} = &\bm{p_e}(\boldsymbol{\delta})-\bm{p_e}(\boldsymbol{\delta}^*)+\epsilon_1 H(\boldsymbol{\delta})C M(\boldsymbol{\omega}-\boldsymbol{\omega}^*),\\
    \frac{\partial V}{\partial \bomega}=&M[\boldsymbol{\omega}-\boldsymbol{\omega}^*+\epsilon_1C (\bm{p_e}(\boldsymbol{\delta})-\bm{p_e}(\boldsymbol{\delta}^*))] -\epsilon_2 M\bm{1}_n\bm{1}_n^\top (\bm{s}-\bm{s}^*),\\
    \frac{\partial V}{\partial \bm{s}}= &C (k\bm{s}-k\bm{s}^*) - \epsilon_2 \bm{1}_n\bm{1}_n^\top M\bomega.
\end{align*}}
Therefore, the partial Lie derivatives can be written as 
 {\small\begin{flalign*}
    \frac{\partial V}{\partial \bdelta}\dot{\bdelta}=&(\bm{p_e}(\boldsymbol{\delta})-\bm{p_e}(\boldsymbol{\delta}^*))^\top(\boldsymbol{\omega}-\bm{1}_n\frac{\bm{1}_n^\top \boldsymbol{\omega}}{n})\\
    &+\epsilon_1 (H(\boldsymbol{\delta})C M(\boldsymbol{\omega}-\boldsymbol{\omega}^*))^\top (\boldsymbol{\omega}-\bm{1}_n\frac{\bm{1}_n^\top \boldsymbol{\omega}}{n}).&&
\end{flalign*}}
\small
\begin{equation*}
    \frac{\partial V}{\partial \bomega}\dot{\bomega} =  [\boldsymbol{\omega}-\boldsymbol{\omega}^*+\epsilon_1 C(\bm{p_e}(\boldsymbol{\delta})-\bm{p_e}(\boldsymbol{\delta}^*))-\epsilon_2 \bm{1}_n\bm{1}_n^\top (\bm{s}-\bm{s}^*)]^\top  M\dot{\bomega}.
\end{equation*}
\normalsize
\small
\begin{subequations}
    \begin{flalign*}
     \frac{\partial V}{\partial \bm{s}}\dot{\bm{s}} &= -(k\bm{s}-k\bm{s}^*)^\top [(\bomega-\bomega^*)+ CEE^\top Ck\bm{s}]\\
    &+\epsilon_2\bomega^\top M\bm{1}_n\bm{1}_n^\top [C ^{-1}(\bomega-\bomega^*)+EE^\top Ck\bm{s}].&&
\end{flalign*}
\end{subequations}
\normalsize

Following Theorem \ref{thm:equilibrium}, $Ck\bm{s}^*\in range(\bm{1}_n)$ and thus $C EE^\top Ck\bm{s}^*=0$. A direct result of this Theorem is
{\small\begin{flalign*}
    &(k\bm{s}-k\bm{s}^*)^\top CEE^\top Ck\bm{s}
    =(k\bm{s}-k\bm{s^*})^\top C EE^\top C (k\bm{s}-k\bm{s^*})&&
\end{flalign*}}

Note that by definition $\bm{p_e}(\boldsymbol{\delta})^\top \bm{1}_n=0,H(\boldsymbol{\delta})^\top \bm{1}_n=0$, $(\bm{p_m}-D\boldsymbol{\omega}^*-\pi(\boldsymbol{\omega}^*)-\bm{p_e}(\boldsymbol{\delta}^*)+k\bm{s}^*=0$ at the equilibrium. In order to simplify the summation, we introduce the following three zero-terms
\begin{subequations}{\footnotesize
    \begin{align*}
        &Z_1= \underbrace{(\bm{p_e}(\boldsymbol{\delta})-\bm{p_e}(\boldsymbol{\delta}^*)+\epsilon_1 H(\boldsymbol{\delta})C M(\boldsymbol{\omega}-\boldsymbol{\omega}^*))^\top (1\frac{1^\top \boldsymbol{\omega}}{n}-1\boldsymbol{\omega}^*)}_{=0}\\
        &Z_2=-(\boldsymbol{\omega}-\boldsymbol{\omega})^\top \underbrace{(\bm{p}-D\boldsymbol{\omega}^*-\pi(\boldsymbol{\omega}^*)-\bm{p_e}(\boldsymbol{\delta}^*)+k\bm{s}^*)}_{=0}\\
        &Z_3=-[\epsilon_1C (\bm{p_e}(\boldsymbol{\delta})-\bm{p_e}(\boldsymbol{\delta}^*))]^\top \underbrace{(\bm{p}-D\boldsymbol{\omega}^*-\pi(\boldsymbol{\omega}^*)-\bm{p_e}(\boldsymbol{\delta}^*)+k\bm{s}^*)}_{=0}
    \end{align*}}
\end{subequations}
Consider $\sip[{Q(\bdelta,\bomega)}]$ as follows
\small
$$\begin{bmatrix}
    \epsilon_1 C & \epsilon_1 C(D+K(\boldsymbol{\omega}))&-\epsilon_1 C\\ 0&D-\epsilon_1 MCH(\boldsymbol{\delta})-\epsilon_2M\bm{1}_n\bm{1}_n^\top C ^{-1}&-\frac{\epsilon_2}{k}(D+K(\boldsymbol{\omega}))\bm{1}_n\bm{1}_n^\top \\
    0&0&C EE^\top C  + \frac{\epsilon_2}{k} \bm{1}_n\bm{1}_n^\top 
\end{bmatrix},$$
\normalsize

Thus, we have
\begin{subequations}{\small
    \begin{align*}
        &\dot{V}(\boldsymbol{\delta},\boldsymbol{\omega},\bm{s})=\frac{\partial V}{\partial \bdelta}\dot{\bdelta}+\frac{\partial V}{\partial \bomega}\dot{\bomega}+\frac{\partial V}{\partial \bm{s}}\dot{\bm{s}}\\
        &=\frac{\partial V}{\partial \bdelta}\dot{\bdelta}+\frac{\partial V}{\partial \bomega}\dot{\bomega}+\frac{\partial V}{\partial \bm{s}}\dot{\bm{s}} + Z_1 + Z_2 + Z_3\\
        &\leq -\begin{bmatrix}
                \bm{p_e}(\boldsymbol{\delta})-\bm{p_e}(\boldsymbol{\delta}^*)\\\boldsymbol{\omega}-\boldsymbol{\omega}^*\\k\bm{s}-k\bm{s^*}
            \end{bmatrix}^\top  Q(\boldsymbol{\delta},\bomega)\begin{bmatrix}
                \bm{p_e}(\boldsymbol{\delta})-\bm{p_e}(\boldsymbol{\delta}^*)\\\boldsymbol{\omega}-\boldsymbol{\omega}^*\\k\bm{s}-k\bm{s^*}
            \end{bmatrix}\\
            &-(\boldsymbol{\omega}-\boldsymbol{\omega}^*)^\top (\pi(\boldsymbol{\omega})-\pi(\boldsymbol{\omega}^*))\\
            &+[\boldsymbol{\omega}-\boldsymbol{\omega}^*+\epsilon_1 C(\bm{p_e}(\boldsymbol{\delta})-\bm{p_e}(\boldsymbol{\delta}^*))-\frac{\epsilon_2}{k} \bm{1}_n\bm{1}_n^\top (k\bm{s}-k\bm{s}^*)]^\top \bm{\Delta d}
    \end{align*}}
\end{subequations}

    Consider a small enough $\epsilon_1$ and $\epsilon_2$, matrix $Q(\boldsymbol{\delta},\bomega)$ can be positive definite \cite{WEITENBERG2018103}. Given that $\pi(\boldsymbol{\omega})$ is monotonically increasing, $-(\boldsymbol{\omega}-\boldsymbol{\omega}^*)^\top (\pi(\boldsymbol{\omega})-\pi(\boldsymbol{\omega}^*))$ is negative. Thus the above equation can be written as 
    \begin{subequations}
        {\small\begin{align*}
            \dot{V}(\boldsymbol{\delta},\boldsymbol{\omega},\bm{s})\leq& -\begin{bmatrix}
                \bm{p_e}(\boldsymbol{\delta})-\bm{p_e}(\boldsymbol{\delta}^*)\\\boldsymbol{\omega}-\boldsymbol{\omega}^*\\k\bm{s}-k\bm{s^*}
            \end{bmatrix}^\top  Q(\boldsymbol{\delta},\bomega)\begin{bmatrix}
                \bm{p_e}(\boldsymbol{\delta})-\bm{p_e}(\boldsymbol{\delta}^*)\\\boldsymbol{\omega}-\boldsymbol{\omega}^*\\k\bm{s}-k\bm{s^*}
            \end{bmatrix}\\
            &+\begin{bmatrix}
                C(\bm{p_e}(\boldsymbol{\delta})-\bm{p_e}(\boldsymbol{\delta}^*))\\\boldsymbol{\omega}-\boldsymbol{\omega}^*\\-\bm{1}_n\bm{1}_n^\top (\bm{s}-\bm{s}^*)
            \end{bmatrix}^\top \begin{bmatrix}
                \epsilon \bm{\Delta d}\\\bm{\Delta d}\\ \epsilon_2 \bm{\Delta d}
            \end{bmatrix}
        \end{align*}}
    \end{subequations}
        
        This further leads to 
        $${\small\dot{V}\leq -\frac{a_3}{a_2}V(\boldsymbol{\delta},\boldsymbol{\omega},\bm{s})+\frac{a_4\sqrt{1+\epsilon^2+(\frac{\epsilon_2}{k})^2}}{\sqrt{a_1}}\lVert \bm{\bm{\Delta d} }\rVert_2 \sqrt{V(\boldsymbol{\delta},\boldsymbol{\omega},\bm{s})}}$$
        where $a_3=\min_{\boldsymbol{\delta,\omega}}\lambda_{\min}(Q(\boldsymbol{\delta},\bomega))\min(1,\eta_1^2)>0$, $a_4=\sqrt{\max(1,\eta_2^2C_{\max}^2,n^2)}>0$. Thus we define $\alpha_1$, $\alpha_2$ as follows,
        \begin{equation}\label{eq:alpha}
             \alpha_1=\frac{a_3}{a_2}, \quad \alpha_2=\frac{a_4\sqrt{1+\epsilon^2+(\frac{\epsilon_2}{k})^2}}{\sqrt{a_1}}.
        \end{equation}
\end{proof}

Given that $\dot{\sqrt{V}}=\frac{\dot V}{2\sqrt{V}}$, using Lemma \ref{lemma:derivative}, we get
$$\dot{\sqrt{V(\boldsymbol{\delta},\boldsymbol{\omega},\bm{s})}}\leq -\frac{\alpha_1}{2}\sqrt{V(\boldsymbol{\delta},\boldsymbol{\omega},\bm{s})} +\frac{\alpha_2}{2}\lVert \bm{\Delta d}\rVert_2.$$
Using the Comparison Lemma \cite[Lemma 3.4]{khalil2002nonlinear}, we have 
\begin{equation*}
\sqrt{V(\boldsymbol{\delta},\boldsymbol{\omega},\bm{s})}\leq \sqrt{V(\boldsymbol{\delta},\boldsymbol{\omega},\bm{s})}|_{t=0}\exp^{-\frac{\alpha_1}{2}t}
        +\frac{\alpha_2}{\alpha_1}\lVert \bm{\Delta d} \rVert_2,
\end{equation*}
where the second exponential term is dropped by relaxation. Utilizing Lemma \ref{lm1}, the above can be rewritten as
\begin{subequations}
    \begin{align}
        &\lVert \bm{x}(t) \rVert_2\leq \kappa \rho^t\lVert \bm{x}_0\rVert_2+\beta\lVert \bm{\Delta d}(t) \rVert_{L_{\infty}}, \\ 
        &\text{where }\kappa =\frac{\sqrt{a_2}}{\sqrt{a_1}},\rho=\exp^{-\frac{\alpha_1}{2}}, \beta = \frac{\alpha_2}{\alpha_1\sqrt{a_1}}.\label{eq:detail_iss} 
    \end{align}
\end{subequations}

Notably, while Exp-ISS is guaranteed regardless of inertia modes, the parameters $\kappa$, $\rho$, and $\beta$ are dependent on the inertia matrix, indicating that variations in inertia can affect the convergence rate.